\RequirePackage{fix-cm}
\documentclass[smallcondensed]{svjour3} %
\usepackage{amsmath}
\usepackage{amssymb}
\usepackage{amsfonts}
\usepackage[T1]{fontenc}
\usepackage{graphicx}%
\usepackage{epsfig}%
\usepackage{calc}%

\newcommand{\pa}[1]{\left( #1 \right)}
\newcommand{\br}[1]{\left[ #1 \right]}
\newcommand{\ac}[1]{\left\{ #1 \right\}}

\newcommand{\ta}{\tau}

\newcommand{\x}{\xi}

\newcommand{\RR}{\mathbb{R}}
\newcommand{\CC}{\mathbb{C}}

\newcommand{\rarrow}{\rightarrow}

\newcommand{\rank}{\operatorname{rank}}

\newcommand{\ie}{\textit{i.e.}\ }

\newcommand{\del}{\partial}

\newcommand{\sech}{\ \mathrm{sech}}

\newenvironment{aleq}{\begin{equation}\begin{aligned}}{\end{aligned}\end{equation}}
\newenvironment{aleq*}{\begin{equation*}\begin{aligned}}{\end{aligned}\end{equation*}}
\newenvironment{gaeq}{\begin{equation}\begin{gathered}}{\end{gathered}\end{equation}}
\newenvironment{gaeq*}{\begin{equation*}\begin{gathered}}{\end{gathered}\end{equation*}}
\newenvironment{eqe}{\begin{equation}}{\end{equation}}

\renewcommand{\thetable}{\Roman{table}}%
\renewcommand{\thesection}{\Roman{section}}%
\renewcommand{\thesubsection}{\Alph{subsection}}%
\renewcommand{\thesubsubsection}{\arabic{subsubsection}}%
\renewcommand{\tablename}{Table}
\begin{document}
%\fontsize{12}{14} \selectfont
\title{Solutions of first-order quasilinear systems expressed in Riemann invariants
}

\titlerunning{Solutions of quasilinear systems.}        % if too long for running head

\author{A.M. Grundland         \and
        V. Lamothe
}


\institute{A.M. Grundland \at
              Centre de Recherches Math\'ematiques, Universit\'e du Montr\'eal, C.P. 6128, Succc.
Centre-ville, Montr\'eal, (QC) H3C 3J7, Canada\\
and D\'epartement de math\'ematiques et informatiques, Universit\'e
du Qu\'ebec, Trois-Rivi\`eres, (QC) G9A 5H7, Canada\\
              \email{grundlan@crm.umontreal.ca}           %  \\
           \and
           V. Lamothe \at
              D\'epartement de Math\'ematiques et Statistique, Universit\'e de Montr\'eal, C.P. 6128, Succc.
Centre-ville, Montr\'eal, (QC) H3C 3J7, Canada\\
            \email{lamothe@crm.umontreal.ca}
 }

\date{Received: date / Accepted: date}

\maketitle

\begin{abstract}
We present a new technique for constructing solutions of quasilinear systems of first-order partial differential equations, in particular inhomogeneous ones. A generalization of the Riemann invariants method to the case of inhomogeneous hyperbolic and elliptic systems is formulated. The algebraization of these systems enables us to construct certain classes of solutions for which the matrix of derivatives of the unknown functions is expressible in terms of special orthogonal matrices. These solutions can be interpreted as nonlinear superpositions of $k$ waves (or $k$ modes) in the case of hyperbolic (or elliptic) systems, respectively. Theoretical considerations are illustrated by several examples of inhomogeneous hydrodynamic-type equations which allow us to construct solitonlike solutions (bump and kinks) and multiwave (mode) solutions.
\end{abstract}
\section{Introduction}
Many nonlinear phenomena appearing in physics are described by first-order quasilinear systems in both their hyperbolic and elliptic regions. These systems have been mostly studied in the case of two independent variables. However, until now, no satisfactory complete theory exists for those systems. For example, with the exception of certain particular results, the existence and uniqueness theorem for solutions of initial and boundary value problems is not available in general. Such exceptions include the necessary and sufficient conditions for the temporal existence of smooth solutions of hyperbolic systems admitting conservation laws \cite{JohnKlainerman:1984,Madja:1984,Rozdestvenski:1983}. Solutions of hyperbolic quasilinear systems do not exist for an arbitrary period of time, even for smooth conditions. They usually blow up at the end of a finite time interval. In general, the first derivatives of the solution become unbounded after a finite time $T>0$, and for a time $t>T$, smooth solutions no longer exist. This well-known phenomenon is called the gradient catastrophe \cite{CourantFriedrich:1958,John:1974,Rozdestvenski:1983}. It can even occur in situations where physical intuition would lead us to expect the existence of continuous solutions after time $T$. In other words, the difficulty which appears here is to determine initial data for the quasilinear hyperbolic system which allows sufficiently large time interval before the gradient catastrophe occurs. It has been proved (see \textit{e.g.} \cite{CourantFriedrich:1958,John:1974,Madja:1984,Rozdestvenski:1983,Whitham:1974}) that even for sufficiently small initial conditions, there exists a time interval $\br{t_0,T}$ in which the gradient catastrophe does not occur. In this interval, the problem of propagation and superposition of waves can be posed and solved by the method of characteristics. Through this method, the existence, uniqueness and continuous dependence of the solution with respect to the initial conditions has been established by many authors (see \textit{e.g.} \cite{Boillat:1965,CourantHilbert:1962,Jeffrey:1976,Lighthill:1968,Mises:1958,Rozdestvenski:1983} and references therein). The obtained results are significant in the sense that the solution is constructed in the domain where its existence is predicted.
\paragraph{}The method of Riemann invariants and its generalization, that is, the generalized method of characteristics (GMC) \cite{Burnat:1969,Jeffrey:1976,Peradzynski:1985} for hyperbolic quasilinear systems of equations in many dimensions, is a technique for obtaining certain classes of exact solutions representing Riemann waves. These solutions are omnipresent for hyperbolic systems and constitute their elementary solutions. They are building blocks for the construction of more general solutions describing the superposition of many waves, $k$-waves (solutions of rank $k$), that are more interesting from the physical point of view \cite{Burnat:1972,GrundlandZelazny:1983,Jeffrey:1976,Peradzynski:1985}. Recently, the applicability of the conditional symmetry method  (CSM) has been extended to the construction of multiwave solutions obtained by the GMC \cite{GrundlandHuard:2006,GrundlandHuard:2007,GrundlandTafel:1996}. The CSM is not limited to hyperbolic systems, but can also be applied to elliptic systems. The links between the two methods is an interesting problem which was addressed in \cite{GrundlandHuard:2007,GrundlandLamothe:2013}. The adaptation of the GMC in the context of the analysis of the symmetry group of first-order quasilinear elliptic systems requires the introduction of complex integral elements instead of simple real integral elements (which are used in the construction of solutions of hyperbolic systems \cite{Burnat:1972,Grundland:1974,Peradzynski:1985,Sobolev:1934}). In particular, for first-order elliptic systems, we are interested in the construction of multimode solutions corresponding to nonlinear superpositions of elementary solutions (simple modes). We demonstrate that the method of conditional symmetries \cite{DoyleGrundland:1996,Fushchych:1991,GrundlandHuard:2007} is an efficient tool for reaching this goal. This approach is applied to hydrodynamic-type equations in their elliptic and hyperbolic regions and has been studied through both the GMC and the CSM. This constitutes the subject of this work, which is a follow up of the paper \cite{GrundlandLamothe:2013}.
\paragraph{}The proposed approach goes deeper into the algebraic aspect of first-order partial differential systems, which allows us to obtain some new results on their solvability. For this purpose we postulate a new form of solution of the initial system for which the matrix of derivatives of the unknown functions is expressed in terms of special orthogonal matrices, of the right-hand side of the partial differential system and of some characteristic vectors associated with the homogeneous part of this system. This decomposition of the matrix of derivatives is no longer in the specific form restricted to the sum of homogeneous and inhomogeneous integral elements as required by the GMC. In the latter case, when solutions exist, they represent a superposition of Riemann waves which admit the freedom of $k$ arbitrary functions of one variable \cite{Cartan:1953,Grundland:1974,Peradzynski:1971}. For the postulate form of the solution the compatibility conditions are weakened in such a way that they allow us to obtain some larger classes of solutions including the one obtained by the GMC for the problem of superposition of waves or modes.
\paragraph{}The plan of this paper is as follows. Section \ref{sec:GMC} contains a brief description of the construction of Riemann $k$ waves obtained by the GMC.  Section \ref{sec:ex:GMC} contains examples of applications of the GMC to the equations of fluid dynamics in $(3+1)$ dimensions. In Section \ref{sec:2}, we investigate and construct the multiwave solutions expressed in terms of Riemann invariants which represent a generalization of the results obtained in Section \ref{sec:GMC}. Section \ref{sec:3} contains a detailed account of the construction of multimode solutions for elliptic systems. Sections \ref{sec:5} and \ref{sec:6} present an adaptation of the method presented in Sections \ref{sec:2} and \ref{sec:3} for the case of underdetermined systems. Several examples of inhomogeneous hydrodynamic-type equations are included in Section \ref{sec:7} as illustrations of the theoretical results. Section \ref{sec:finalremarks} contains remarks and some suggestions regarding possible future developments.
%#################### Method of characteristics
\section{Generalized method of characteristics}\label{sec:GMC}
Consider a first-order system of quasilinear autonomous partial differential equations (PDEs) in $p$ independent variables
\begin{eqe}\label{eq:omr:1}
\mathcal{A}^{\mu i}_\alpha(u) u^\alpha_i=b^\mu(u),\quad \mu=1,\ldots,m,\quad \alpha=1,\ldots,q,\quad i=1,\ldots,p,
\end{eqe}%
where $u^\alpha_i=\del u^\alpha/\del x^i$. We adopt the convention that repeated indices are summed unless one of them is in brackets. The system (\ref{eq:omr:1}) is an inhomogeneous one with coefficients depending on $q$ unknown functions $u=(u^1,\ldots,u^q)^T\in \RR^q$. The Euclidean space $E=\RR^p$ (the space of independent variables $x=(x^1,\ldots,x^p)$) is called the physical space, while the space $\mathcal{U}=\RR^q$ (the space of dependent variables) is called the hodograph space. Let us assume that the coefficients $A^{\mu i}_\alpha(u)$ and $b(u)$ are smooth real-valued functions of the real variables $u$. We investigate the existence and the construction of solutions describing the propagation and nonlinear superposition of waves or modes that can be admitted by the system (\ref{eq:omr:1}). Such solutions are particularly interesting from the physical point of view because they cover a wide range of the wave phenomena arising in the presence of the external forces that can be observed in such domains as field theory, fluid dynamics, elasticity, \textit{etc}. These phenomena are described by systems of the form (\ref{eq:omr:1}) or systems that can be reduced to that form by introducing new unknown functions. The methodological approach assumed in this section is based on the generalized method of characteristics which was extensively developed in \cite{Boillat:1965,Burnat:1969,Jeffrey:1976,Peradzynski:1985} for homogeneous systems and generalized to inhomogeneous systems \cite{GrundlandZelazny:1983,Grundland:1974}. The specific feature of this approach is an algebraic and geometric point of view. The algebraization of PDEs was made possible by representing the general integral element as a linear combination of some special simple integral elements \cite{Burnat:1969,GrundlandZelazny:1983,Peradzynski:1985}.
%##################
\paragraph{}The algebraization of the inhomogeneous system (\ref{eq:omr:1}) allows us to construct certain classes of solutions, which correspond to a superposition of $k$ simple waves together with a simple state solution (as introduced in \cite{GrundlandZelazny:1983}, see Appendix). In this case, a specific constraint on the derivatives of a solution $u(x)$ is postulated. We require that the Jacobian matrix $\del u^\alpha/\del x^i$ be the sum of homogeneous and inhomogeneous simple integral elements
\begin{eqe}\label{eq:GMC:2}
\frac{\del u^\alpha}{\del x^i}=\xi^A\gamma_A^\alpha\lambda^A_i+\gamma^\alpha_0 \lambda^0_i,
\end{eqe}%
where the summation is taken over the index $A$ which runs from 1 to $k$ and we have assumed that
\begin{gaeq}\label{eq:GMC:ast}
\mathcal{A}_\alpha^{\mu i}\gamma^\alpha_{(A)}\lambda_i^A=0,\qquad \mathcal{A}_\alpha^{\mu i}\gamma_{0}^\alpha\lambda^0_i=b^\mu,\\
\lambda^{A_1}\wedge\lambda^{A_2}\wedge\lambda^{A_3}\neq 0\quad\text{for }0<A_1<A_2<A_3\leq k,
\end{gaeq}%
and
$$\rank\pa{\gamma_0,\gamma_1,\ldots,\gamma_k}=k+1\leq q.$$%
Here $\xi^A\neq0$ is treated as a function of $x$. We suppose that we can find $k$ distinct characteristic vectors $\lambda^A(u)=\pa{\lambda^A_1(u),\ldots, \lambda^A_p(u)}$ and $k$ linearly independent characteristic vectors $\gamma_A(u)=\pa{\gamma_A^1(u),\ldots,\gamma_A^q(u)}$ which satisfy the homogenous equations (\ref{eq:GMC:ast}). We suppose also that the noncharacteristic vectors $\lambda^0(u)=\pa{\lambda^0_1(u),\ldots,\lambda^0_p(u)}$ and $\gamma_0(u)=\pa{\gamma_0^1(u), \ldots, \gamma_0^q(u)}$ obey the inhomogeneous equations (\ref{eq:GMC:ast}). The resulting solution of (\ref{eq:GMC:2}) is called a $k$-wave solution for the inhomogeneous hyperbolic system \cite{GrundlandZelazny:1983,Grundland:1974}. Under the above assumptions it was shown \cite{GrundlandVassiliou:1991,GrundlandZelazny:1983,Peradzynski:1985} that the necessary and sufficient conditions for the existence of $k$-wave solutions of the system (\ref{eq:omr:1}), subjected to conditions (\ref{eq:GMC:2}), require the following constraints on vectors $\lambda^n$ and $\gamma^n$, $n=0,1,\ldots,k$. The explicit parametrization of the integral surface $S$ in terms of Riemann invariants $r=(r^0,r^1,\ldots, r^k)$ is obtained by solving the system of PDEs
\begin{eqe}\label{eq:GMC:3}
\frac{\del f^\alpha}{\del r^n}=\gamma^\alpha_n(u),\qquad n=0,1,\ldots,k,
\end{eqe}%
with solution
%\begin{eqe}\label{eq:GMC:4}
$$u=f(r).$$
%\end{eqe}%
Next, we look for the most general solution of the system of 1-forms $dr^n$ for $\lambda^n$ as functions of $r$
\begin{eqe}\label{eq:GMC:dr}
dr^0=\lambda^0,\qquad dr^A=\xi^A(x)\lambda^A(r),\qquad A=1,\ldots,k.
\end{eqe}%
The compatibility conditions for the system (\ref{eq:GMC:dr}) impose some restrictions on the wave vectors $\lambda^n$. Namely, the system (\ref{eq:GMC:dr}) has solutions (is completely integrable) if the following conditions are satisfied
\begin{eqe}\label{eq:GMC:5}
\frac{\del \lambda^A}{\del r^n}\in\operatorname{span}\ac{\lambda^A,\lambda^n},\qquad \frac{\del \lambda^0}{\del r^n}\in\operatorname{span}\ac{\lambda^n},\qquad n\neq A.
\end{eqe}%
Here we have denoted the vector space spanned by the vectors $\lambda^A$ and $\lambda^n$ by $\operatorname{span}\ac{\lambda^A,\lambda^n}$, while $\del \lambda^0/ \del r^n$ is proportional to $\lambda^n$.
These relations are the necessary and sufficient conditions for the existence of solutions of the system (\ref{eq:GMC:dr}). They ensure that the set of solutions of the system (\ref{eq:GMC:dr}) depends on $k$ arbitrary functions of one variable \cite{GrundlandVassiliou:1991,GrundlandZelazny:1983}.
Finally, under the assumption that $\lambda^1,\ldots,\lambda^k$ are linearly independent and satisfy (\ref{eq:GMC:5}) the $k$-wave solutions are obtained from the explicit parametrization of the integral surface $u=f(r^0,r^1,\ldots,r^k)$. The quantities $r^0,r^1,\ldots, r^k$ are implicitly defined as functions of $x^1,\ldots, x^p$ by the solution of the system for certain functionally independent differentiable functions $\psi^A$ of $r$
\begin{eqe}%\label{eq:GMC:6}
\lambda_i^A(r)x^i=\psi^A(r),\qquad \frac{\del \psi^A}{\del r^n}=\alpha_n^A(r)\psi^{(A)}+\beta^A_n(r)\psi^{(n)},\quad A\neq n,
\end{eqe}%
where $\alpha^A_n$ and $\beta^A_n$ are given functions of $r=(r^0,r^1,\ldots,r^k)$. The physical interpretation of these solutions is that the profiles of simple waves related with the simple elements $\gamma_A\otimes\lambda^A$ can be chosen in an arbitrary way, but the profile of a simple state related with the inhomogeneous element $\gamma_0\otimes \lambda^0$ is somehow determined by (\ref{eq:GMC:ast}) and (\ref{eq:GMC:5}). These solutions represent some nonlinear superpositions of $k$ waves on a simple state \cite{GrundlandZelazny:1983}. However, our present approach will go deeper into the geometrical aspects of compatibility conditions (\ref{eq:GMC:3}) and (\ref{eq:GMC:5}) by weakening them. So it will later enable us to obtain new results on the solvability of the problem of superposition of $k$-waves for inhomogeneous hyperbolic systems and extend this approach to elliptic systems.
%########
%########
\paragraph{}Note that, in the case of the underdetermined system of equations (\ref{eq:omr:1}) (\ie when $m<q=m+l$) the real simple integral elements are determined by the equation
\begin{eqe}%\label{eq:GMC:star}
\mathcal{A}_\alpha^{\mu i}(u) \lambda_i \eta^\alpha=\mathcal{A}_s^{\mu i}(u)\lambda_i\eta_1^s+\mathcal{A}_t^{\mu i}(u)\lambda_i \eta_2^t=b^\mu,
\end{eqe}%
where $\mu\in M=\ac{1,\ldots,m}$, $s\in I_1$, $t\in I_2$ and $I_1\cup I_2$ is the partition of the set $I=\ac{1,\ldots,q}$ into  subsets containing $m$ and $l=q-m$ elements, respectively. The vector field $\eta=(\eta_1,\eta_2)$ has $q$ components, while
$$\eta_1=\pa{\eta^s}_{s\in I_1},\qquad \eta_2=(\eta^t)_{t\in I_2}$$%
and the $m\times q$ matrix
$$\pa{\mathcal{A}_1\lambda, \mathcal{A}_2\lambda}=\pa{\pa{\mathcal{A}_s^{\mu i}\lambda_i}^{\mu\in m}_{s\in I_1},\pa{\mathcal{A}_t^{\mu i}\lambda_i}^{\mu\in m}_{t\in I_2}}$$%
is a proper partition of the matrix $\pa{\mathcal{A}_\alpha^{\mu i}\lambda_i}$. If the $m\times m$ matrix $(\mathcal{A}_1\lambda)$ is a nonsingular one, then we can determine the quantity
$$\eta_1=(\mathcal{A}_1\lambda)^{-1}\pa{b-\pa{\mathcal{A}_2\lambda}\eta_2}.$$%
Hence, there exists a bijective correspondence between $\eta\otimes \lambda$ and $\eta_2\otimes \lambda$ which determines a map on the set of simple integral elements and for which the domain is formed by these elements that have nonsingular matrices
$$\mathcal{A}_1\lambda=\pa{\mathcal{A}_s^{\mu i}\lambda_i}_{s\in I_1}^{\mu\in m}.$$%
This map is determined by the partition of the set $I=\ac{1,\ldots,q}=I_1\cup I_2$. All possible partitions of this type determine an atlas composed of $\pa{\begin{array}{c}q\\ m\end{array}}$ maps covering the set of real simple integral elements, where they are regular in the sense that
$$\rank\pa{\mathcal{A}_\alpha^{\mu i}\lambda_i}=m.$$%
One can partition the set of simple integral elements in stratas numbered by the rank of matrices $\mathcal{A}_\alpha^{\mu i}\lambda_i$. Regular elements form strata of the highest dimension. One can determine atlases on the other stratas analogously and solve the considered underdetermined system by the GMC.
%################### Examples - GMC #################
\section{Inhomogeneous fluid dynamics equations}\label{sec:ex:GMC}
Now we present some examples which illustrate the theoretical considerations presented in Section \ref{sec:GMC}. We discuss the classical equations of an ideal compressible nonviscous fluid placed in the presence of gravitational and Coriolis forces. Under these assumptions, the fluid dynamics system of equations in $(3+1)$ dimensions takes the form
\begin{aleq}\label{ex:GMC:1}
&\rho\pa{\vec{v}_t+(\vec{v}\cdot\nabla)\vec{v}}+\nabla p=\rho \vec{v}\times\vec{\Omega}+\rho \vec{g},\\
&\rho_t+(\vec{v}\cdot \nabla)\rho+\rho \nabla\cdot\vec{v}=0,\\
&(p \rho^{-k})_t+(\vec{v}\cdot \nabla)(p\rho^{-k})=0,
\end{aleq}%
where we have used the following notation: $\rho$ and $p$ are the density and the pressure of the fluid respectively, $\vec{v}$ is the vector field of the fluid velocity, $k>0$ is the polytropic exponent and the external forces are the Coriolis $\rho \vec{v}\times\vec{\Omega}$ (with angular velocity $\vec{\vec{\Omega}}$) and the gravitational $\rho \vec{g}$ forces. The algebraic equations that determine simple integral elements for the equation (\ref{ex:GMC:1}) are of the form
\begin{aleq}\label{ex:GMC:2}
&\rho \delta|\vec{\lambda}| \vec{\gamma}+\gamma^p\vec{\lambda}=\rho \vec{v}\times\vec{\Omega}+\rho \vec{g},\\
&\delta|\vec{\lambda}| \gamma^\rho+\rho \vec{\gamma}\cdot \vec{\lambda}=0,\\
&\rho\delta|\vec{\lambda}|(\gamma^p-\frac{k p}{\rho}\gamma^\rho)=0.
\end{aleq}%
Here we have used the following notation: The space of unknown functions, the hodograph space $\mathcal{U}\subset \RR^5$, has coordinates $u=(\rho,p,\vec{v})$. The corresponding elements of the tangent space $T_{u}\mathcal{U}$ are denoted by $\gamma=(\gamma^\rho,\gamma^p,\vec{\gamma})$, where $\vec{\gamma}=(\gamma^1,\gamma^2,\gamma^3)$ is associated with the velocity vector $\vec{v}$ and $(\gamma^\rho,\gamma^p)$ are associated with the density $\rho$ and pressure $p$, respectively. Here we have denoted $\lambda=(\lambda_0,\vec{\lambda})$, where $\lambda_0$ is the phase velocity and $\vec{\lambda}=(\lambda_1,\lambda_2,\lambda_3)$ is the wave vector. We have replaced the derivatives of the unknown functions $\rho_t$, $\rho_{x^i}$, $p_t$, $p_{x^i}$, $\vec{v}_t$ and $\vec{v}_{x^i}$ in equation (\ref{ex:GMC:1}) by the simple elements $\gamma^\rho \lambda_0$, $\gamma^\rho\lambda_i$, $\gamma^p\lambda_0$, $\gamma^p\lambda_i$, $\vec{\gamma}\lambda_0$ and $\vec{\gamma}\lambda_{i}$, $i=1,2,3,$ respectively. According to \cite{Peradzynski:1985}, we define the quantity
\begin{eqe}\label{ex:GMC:ast}
\delta |\vec{\lambda}|=\lambda_0+\vec{v}\cdot \vec{\lambda},
\end{eqe}%
which physically describes the velocity of propagation of a wave relative to the fluid while $\lambda_0$ describes the phase velocity of the considered wave $\mathfrak{E}$, $\mathfrak{A}$ or $\mathfrak{H}$. The equations (\ref{ex:GMC:2}) form a system of linear inhomogeneous algebraic equations, whose solutions $\gamma\in T_u(\mathcal{U})$ and $\lambda\in \mathfrak{E}^\ast$ ($\mathfrak{E}^\ast$ being the dual space of the classical space-time $\mathfrak{E}\in\RR^4$), determine the simple integral elements associated with the equation (\ref{ex:GMC:1}).
\paragraph{}It follows from the analysis of the homogeneous system corresponding to (\ref{ex:GMC:2}) that there exist nontrivial solutions for the vector $\gamma=(\gamma^\rho,\gamma^p,\vec{\gamma})$ when the characteristic determinant of this system is equal to zero. Thus we obtain the following condition
\begin{eqe}%\label{ex:GMC:3}
\delta^3|\vec{\lambda}|^3\pa{\delta^2|\vec{\lambda}|^2-\frac{k p}{\rho}|\vec{\lambda}|^2}=0.
\end{eqe}%
So we obtain two types of  homogeneous simple integral elements
\begin{itemize}
\item[(\textbf{i}) ] The entropic simple element $\mathfrak{E}$:
\begin{eqe}\label{ex:GMC:4}
\delta=0,\qquad \lambda=(-\vec{v}\cdot\vec{\lambda}, \vec{\lambda}),\qquad \gamma=(\gamma^\rho,0,\vec{\gamma}),
\end{eqe}%
with the condition $\vec{\gamma}\cdot \vec{\lambda}=0$. Here $\gamma_\rho$, is an arbitrary function.
\item[(\textbf{ii})] The acoustic simple element $\mathfrak{A}$:
\begin{aleq*}%\label{ex:GMC:5}
\delta|\vec{\lambda}|=&\epsilon\pa{\frac{k p}{\rho}}^{1/2}|\vec{\lambda}|,\quad \lambda=(\delta|\vec{\lambda}|-\vec{v}\cdot\vec{\lambda},\vec{\lambda}),\quad \epsilon=\pm 1,\\ \gamma=&\pa{\gamma^\rho,\frac{k p}{\rho}\gamma^\rho,-\delta\frac{\vec{\lambda}}{|\vec{\lambda}|}\frac{\gamma^\rho}{\rho}},
\end{aleq*}%
where $\gamma^\rho$ is an arbitrary function.
\end{itemize}
The nontrivial solutions for $\gamma_0\in T_u\mathcal{U}$ and $\lambda^0\in E^\ast$ of the inhomogeneous system (\ref{ex:GMC:2}) determine three types of the inhomogeneous simple integral elements of (\ref{ex:GMC:1}), namely
\begin{itemize}
\item[(\textbf{i})  ] The entropic inhomogeneous simple element $\mathfrak{E}_0$
\begin{eqe}\label{ex:GMC:6}
\delta=0,\quad \gamma_0^{\mathfrak{E}_0}=(\gamma_\rho,\rho,\vec{\alpha}\times(\vec{g}-\vec{\Omega}\times \vec{v})),\quad \lambda^0=(-\vec{v}\cdot\vec{g},\vec{g}-\vec{\Omega}\times \vec{v}),
\end{eqe}%
where $\vec{\alpha}$ is an arbitrary vector in $\RR^3$ and $\gamma_\rho$ is an arbitrary function.
\item[(\textbf{ii}) ] The acoustic inhomogeneous simple element $\mathfrak{A}_0$
\begin{gaeq}\label{ex:GMC:7}
\delta|\vec{\lambda}|=\epsilon|\vec{\lambda}|\pa{\frac{k p}{\rho}}^{1/2},\quad \lambda^0_{\mathfrak{A}_0}=\pa{\delta |\vec{\lambda}|-\vec{v}\cdot\vec{\lambda},\vec{\lambda}},\quad \vec{\lambda}\cdot (\vec{g}-\vec{\Omega}\times \vec{v})=0,\\
\gamma_0^{\mathfrak{A}_0}=\pa{\gamma_\rho,\frac{k p}{\rho}\gamma_\rho,\pa{\delta |\vec{\lambda}|}^{-1}\pa{\rho(\vec{g}-\vec{\Omega}\times \vec{v})-\frac{k p}{\rho}\gamma_\rho\vec{\lambda}}},\quad \epsilon=\pm 1.
\end{gaeq}%
where $\gamma_\rho$ is an arbitrary function.
\item[(\textbf{iii})] The hydrodynamic inhomogeneous simple element $\mathfrak{H}_0$
\begin{aleq*}%\label{ex:GMC:8}
\gamma_0^\mathfrak{H}=&\bigg(-\rho\frac{(\vec{g}-\vec{\Omega}\times \vec{v})\cdot \vec{\lambda}}{(\delta|\vec{\lambda}|)^2-\frac{k p}{\rho}|\lambda|^2},-k p \frac{(\vec{g}-\vec{\Omega}\times \vec{v})\cdot \vec{\lambda}}{(\delta|\vec{\lambda}|)^2-\frac{k p}{\rho}|\lambda|^2},\\
&\quad\frac{1}{\rho \delta}\br{\rho (\vec{g}-\vec{\Omega}\times \vec{v})+k p \frac{(\vec{g}-\vec{\Omega}\times \vec{v})\cdot \vec{\lambda}}{(\delta|\vec{\lambda}|)^2-\frac{k p}{\rho}|\lambda|^2 }\vec{\lambda}}\bigg),\\
\lambda=&(\delta |\vec{\lambda}|-\vec{v}\cdot\vec{\lambda},\vec{\lambda}).
\end{aleq*}%
\end{itemize}
Here $\delta|\vec{\lambda}|$ is different from 0 and $\epsilon\pa{{k p}/{\rho}}^{1/2}$, $\epsilon=\pm 1$, and otherwise arbitrary. From the definition (\ref{ex:GMC:ast}), we obtain that the velocity associated with the solution of $\mathfrak{E}_0$ moves together with the fluid. For the case of acoustic solutions $\mathfrak{A}_0$ the velocity of the fluid is equal to the sound velocity: ${d p}/{d\rho}=\pa{{k p}/{\rho}}^{1/2}$, while the hydrodynamic solution $\mathfrak{H}_0$ moves with any velocity other than the entropic velocity $\delta=0$ or acoustic velocity $\delta|\vec{\lambda}|=\epsilon\pa{{k p}/{\rho}}^{1/2}|\vec{\lambda}|$.
\paragraph{}We now consider the classes of solutions $u$ for which the matrix of the tangent mapping $\del u^\alpha/\del x^{i}$ is the sum of one homogeneous and one inhomogeneous simple integral elements (when $k=1$ in equation (\ref{eq:GMC:2}))
\begin{eqe}\label{ex:GMC:9}
\frac{\del u^\alpha}{\del x^{i}}=\xi\gamma_1^\alpha\lambda_i^1+\gamma_0^\alpha \lambda_i^0,\quad \lambda^1\wedge \lambda^0\neq 0,
\end{eqe}%
where $\xi\not\equiv0$ is treated as an arbitrary function of $x$. According to \cite{GrundlandZelazny:1983} the necessary and sufficient conditions for the existence of solutions of the system (\ref{ex:GMC:9}) require that the commutator of the vector fields $\gamma_0$ and $\gamma_1$ be a linear combination of these fields
%\begin{eqe}\label{ex:GMC:10}
$$
\br{\gamma_0,\gamma_1}\in\operatorname{span}\ac{\gamma_0,\gamma_1}.
$$%\end{eqe}%
This means that $\gamma_0$ and $\gamma_1$ constitute a holonomic system in the sense that there exists a parametrization of a surface
%\begin{eqe}\label{ex:GMC:11}
$$
u=f(r^0,r^1),
$$%\end{eqe}%
tangent to the vector field $\gamma_0$ and $\gamma_1$ such that
\begin{eqe}\label{ex:GMC:12}
\frac{\del f(r^0,r^1)}{\del r^0}=\gamma_0\pa{f(r^0,r^1)},\quad \frac{\del f(r^0,r^1)}{\del r^1}=\gamma_1(f(r^0,r^1))
\end{eqe}%
holds. Consequently the system (\ref{ex:GMC:9}) together with the assumption that $\gamma_0$ and $\gamma_1$ are linearly independent, require us to solve the following system of PDEs
\begin{eqe}\label{ex:GMC:13}
\frac{\del r^0}{\del x^{i}}=\lambda_i^0(r^0,r^1),\quad \frac{\del r^1}{\del x^{i}}=\xi\lambda_i^1(r^0,r^1).
\end{eqe}%
The involutivity condition for the system (\ref{ex:GMC:13}) has already been investigated in \cite{GrundlandZelazny:1983}. These conditions lead to restrictions on the wave vectors $\lambda^0$ and $\lambda^1$. It was shown that the system (\ref{ex:GMC:13}) is completely integrable if the following conditions are satisfied
\begin{eqe}\label{ex:GMC:14}
\frac{\del \lambda^0}{\del r^0}=\alpha_0 \lambda^0,\qquad \frac{\del \lambda^0}{\del r^1}=\alpha_1 \lambda^1,\qquad \frac{\del \lambda^1}{\del r^0}=\beta_0 \lambda^0+\beta_1\lambda^1,\qquad \lambda^0\wedge \lambda^1\neq 0,
\end{eqe}%
where the coefficients $\alpha_0$, $\alpha_1$, $\beta_0$ and $\beta_1$ are functions of $r^0$ and $r^1$ to be determined. Consider the following two cases:
\paragraph{\textbf{1}.}When $\alpha_1\neq 0$, the equations (\ref{ex:GMC:14}) can be integrated and the wave vectors $\lambda^1$ and $\lambda^0$ take the form
%\begin{eqe}\label{ex:GMC:17}
$$
\lambda^0=\lambda(r^1)\exp\varphi(r^0,r^1),\qquad \lambda^1=\frac{\del \varphi}{\del r^1}\lambda(r^1)+\frac{d\lambda}{d r^1},
$$%\end{eqe}%
where the function $\varphi$ is a solution of the equation
%\begin{eqe}\label{ex:GMC:16}
$$
\alpha_0=\frac{\del \varphi}{\del r^0}
$$%\end{eqe}%
for a given function $\alpha_0$ of $r^0$ and $r^1$. All solutions of the system (\ref{ex:GMC:13}) can be obtained by solving the implicit relation with respect to the variables $r^0$, $r^1$ and $x^i$
\begin{aleq}\label{ex:GMC:15}
\lambda(r^1)x^i=&\phi(r^1)+\int_0^{r^0}\exp\pa{-\varphi(s,r^1)}ds,\\
\frac{d\lambda_i(r^1)}{d r^1}x^i=&\dot{\phi}(r^1)-\int_0^{r^0}\frac{\del \varphi(s,r^1)}{\del r^1}\exp\pa{-\varphi(s,r^1)}ds,
\end{aleq}%
\paragraph{\textbf{2}.}When $\alpha_1=0$, the equations (\ref{ex:GMC:14}) can be integrated and the wave vectors $\lambda^1$ and $\lambda^0$ are
%\begin{eqe}\label{ex:GMC:17a}
$$
\lambda^0=ce^{\phi(r^0)},\quad \lambda^1= \chi(r^0,r^1)C+a(r^1),
$$%\end{eqe}%
where $\phi$ and $a$ are arbitrary functions of their arguments, $\chi$ is an arbitrary function of $(r^0,r^1)$ and $C=(c_0,\vec{c})\in \RR^4$ is a constant vector and where $\vec{c}=(c_1,c_2,c_3)$. The general integral of the system (\ref{ex:GMC:13}) can be obtained by solving the implicitly defined relations between the variables $r^0,r^1$ and $x^i$
\begin{gaeq}\label{ex:GMC:17b}
c_ix^i+c_0=\int_0^{r^0}\exp(-\phi(s))ds,\\
\int_0^{r^0}\chi(s,r^1)\exp(-\phi(s))ds+a_i(r^1)x^i=\psi(r^1),
\end{gaeq}%
where $\psi$ and $a_i$ are arbitrary functions of $r^1$. If (\ref{ex:GMC:17b}) can be solved, so that $r^0$, $r^1$ and $u^\alpha$ can be given as a graph over an open set $\mathcal{D}\subset \RR^4$, then the functions $u^\alpha=f^\alpha(r^0,r^1)$, determined from (\ref{ex:GMC:12}), constitute an exact solution (written in terms of the Riemann invariants $r^0$, $r^1$) of the inhomogeneous system (\ref{ex:GMC:9}).  A similar statement holds for the case $\alpha_1\neq 0$ and when one replaces the system (\ref{ex:GMC:17b}) by the system (\ref{ex:GMC:15}).
\paragraph{}Superpositions of simple waves and a simple state, which are solutions of the system (\ref{ex:GMC:1}) and subjected to the differential constraints (\ref{ex:GMC:9}) are illustrated in Table I,
\begin{table}
\begin{center}
\begin{tabular}{|c|c|c|c|}
  \hline
  simple wave$\backslash$ simple state & \ $\mathfrak{E}_0$\  & $\ \mathfrak{A}_0\ $ & $\ \mathfrak{H}_0\ $ \\
\hline
  $\mathfrak{E}$ & $+$ & $+$ & $+$ \\
\hline
  $\mathfrak{A}$ & $+$ & $-$ & $+$ \\
  \hline
\end{tabular}
\end{center}
\caption{The results of superpositions of simple waves determined by (\ref{ex:GMC:1}) and  (\ref{ex:GMC:9})} are summarized as follows.
\end{table}
where $+$ denotes that there exists a superposition written in terms of Riemann invariants and $-$ denotes that there is no superposition. For convenience, we denote by $\mathfrak{EE}_0$, $\mathfrak{EA}_0$, $\textit{etc.}$, the solutions which result from nonlinear superpositions of waves and states associated with the given wave vectors $\lambda$ and $\lambda^0$ given by (\ref{ex:GMC:4}) and (\ref{ex:GMC:6}) or $\lambda$ and $\lambda^0$ given by (\ref{ex:GMC:4}) and (\ref{ex:GMC:7}), respectively. We give several examples to illustrate the construction introduced in this section.
\paragraph{\textbf{1}. The entropic simple wave and the entropic state, $\mathfrak{EE}_0$.} A solution exists, provided that $\vec{g}=(0,0,g)$, $|\vec{\Omega}|=1$ and $\vec{g}\cdot \vec{\Omega}=0$, and it is given by
\begin{gaeq*}%\label{ex:GMC:18}
\rho=\dot{p}(r^0),\qquad p=p(r^0),\\
\vec{v}=-e^{\phi_0}|\vec{g}|^{-2}c_0 \vec{g}+v_2(r)\vec{\Omega}+\pa{\mp e^{\phi_0}|\vec{g}|^{-2}(|\vec{g}|^2-c_0^2)^{1/2}+1}\vec{g}\times\vec{\Omega}
\end{gaeq*}%
with Riemann invariants
\begin{gaeq*}%\label{ex:GMC:19}
c_0 t\pm |\vec{g}|^{-2}(|\vec{g}|^2-c_0^2)^{1/2}\vec{g}\cdot \vec{x}-c_0|\vec{g}|^{-2}(\vec{g}\times\vec{\Omega})\cdot\vec{x}+c_1=e^{-\phi_0}r^0,\\
e^{-\phi_0}\int_0^{r^0}\chi(r,r^1)dr+\br{e^{\phi_0}(\pm( |\vec{g}|^2-c_0^2)^{1/2}a_3+c_0a_1)-a_3|\vec{g}|^2}t\\
+a_1\vec{g}\cdot\vec{x}+a_3(\vec{g}\times\vec{\Omega})\cdot\vec{x}=\psi(r^1),
\end{gaeq*}%
where $c_0=g (1+g^4)^{-1}$, $e^{\phi_0}=c_0^{-1}$, $\phi_0$ and $c_1$ are arbitrary constants and $p(r^0) (>0)$, $v_2(r)$, $\chi(r)$, $a_1(r^1)$, $a_3(r^1)$ and $\psi(r^1)$ are arbitrary functions satisfying $\dot{p}>0$, ${\del v_2}/{\del r^1}\neq 0$ and $\pm(|\vec{g}|^2-c_0^2)^{1/2}a_3\neq -c_0a_1$.
\paragraph{}The following additional entropic solution exists only if $\vec{g}\cdot \vec{\Omega}\neq 0$. It is given by
\begin{gaeq*}%\label{ex:GMC:20}
\vec{g}=(0,0,g),\quad \vec{\Omega}=(0,0,1),\quad \rho=\dot{p}(r^0),\quad p=p(r^0),\\
\vec{v}(r^1)=v_1(r^1)\vec{g}+\bigg( \frac{-|\vec{g}|^2}{(\vec{g}\cdot\vec{\Omega})}\int_0^{r^1}\br{\dot{v}_1(r)(1-v_3(r))+\dot{v}_3(r)v_1(r)}dr\\
-(\vec{g}\cdot\vec{\Omega})\int_0^{r^1}\br{\dot{v}_1(r) v_3(r)-v_1(r)\dot{v}_3(r)}dr+c\bigg)\vec{\Omega}+v_3(r^1)\vec{g}\times\vec{\Omega},
\end{gaeq*}%
where the Riemann invariants are given in implicit form by
\begin{gaeq*}%\label{ex:GMC:21}
\bigg(-v_1|\vec{g}|^2+|\vec{g}|^2\int_0^{r^1}\br{\dot{v}_1(1-v_3)+\dot{v}_3v_1}dr+(\vec{g}\cdot\vec{\Omega})^2\int_0^{r^1}\br{\dot{v}_1v_3-v_1\dot{v}_3}dr \\ -c(\vec{g}\cdot\vec{\Omega})\bigg)t+\pa{(1-v_3)\vec{g}+v_3(\vec{g}\cdot\vec{\Omega})\vec{\Omega}+v_1\vec{g}\times\vec{\Omega}}\cdot\vec{x}=\phi(r^1)+r^0,\\
\pa{(\vec{g}\cdot\vec{\Omega})^2-|\vec{g}|^2}(\dot{v}_1v_3-v_1\dot{v}_3)t+\bigg(-\dot{v}_3\vec{g}\\
+\dot{v}_3(\vec{g}\cdot\vec{\Omega})\vec{\Omega}+\dot{v}_1\vec{g}\times\vec{\Omega}\bigg)\cdot\vec{x}=\dot{\phi}(r^1),
\end{gaeq*}%
where $c$ is an arbitrary constant, $p(r^0)$ $(>0)$, $v_1(r^1)$, $v_3(r^1)$ and $\phi(r^1)$ are arbitrary functions such that $\dot{p}(r^0)>0$ and $\dot{v}_1(r^1)\neq0$ or $\dot{v}_3(r^1)\neq0$.
%####
\paragraph{\textbf{2}. The simple entropic wave and the simple acoustic state, $\mathfrak{EA}_0$.} In the case when $\vec{\lambda}^0=\epsilon_1|\vec{\lambda}|^0\vec{\Omega}$ and $\vec{g}\cdot\vec{\Omega}=0$, the solution is given by
\begin{gaeq*}%\label{ex:GMC:22}
\vec{g}=(0,0,g),\quad \vec{\Omega}=(\Omega_1,\Omega_2,0),\ |\vec{\Omega}|=1,\\
\rho=\rho_0,\qquad p=p_0,\\
\vec{v}=b_2(r^1)\cos\pa{\epsilon \pa{\frac{\kappa p_0}{\rho_0}}^{-1/2}\int_0^{r^0}e^{-\phi(r)}dr+b_1(r^1)}\vec{g}+B_0\vec{\Omega}\qquad\qquad\qquad\\
+\br{b_2(r^1)\sin\pa{\epsilon\pa{\frac{\kappa p_0}{\rho_0}}^{-1/2}\int_0^{r^0}e^{-\phi(r)}dr+b_1(r^1)}+1}\vec{g}\times\vec{\Omega},
\end{gaeq*}%
where the Riemann invariants are
\begin{gaeq*}%\label{ex:GMC:23}
\pa{\epsilon\pa{\frac{\kappa p_0}{\rho_0}}^{1/2}-\epsilon_1 B_0}t+\epsilon_1\vec{\Omega}\cdot\vec{x}+c_1=\int_0^{r^0}e^{-\phi(r)}dr,\\
B_0t-\vec{\Omega}\cdot\vec{x}=\psi(r^1),
\end{gaeq*}%
 where $\rho_0$ $(>0)$, $p_0$ $(>0)$, $B_0$ and $c_1$ are arbitrary constants, $\epsilon_1=\pm 1$, and $b_1(r^1)$, $b_2(r^1)\neq 0$, $\phi(r^0)$ and $\psi(r^1)$ are arbitrary functions such that $\dot{b}_2(r^1)\neq 0$.
 %######
 \paragraph{\textbf{3}. The simple entropic wave and the simple hydrodynamic state, $\mathfrak{EH}_0$.} When $\vec{\lambda}^0\times \vec{\Omega}\neq0$, $\vec{\lambda}^0\cdot\vec{\Omega}=0$, $\dot{p}(r^0)=0$ and $\vec{g}\cdot\vec{\Omega}\neq 0$,  the solution is given by
\begin{gaeq*}%\label{ex:GMC:24}
\vec{g}=(0,0,g),\quad \vec{\Omega}=(0,0,1),\\
\rho=\rho(r^1
),\qquad p=p_0,\\
\vec{v}=-\vec{g}\cdot(\vec{c}\times \vec{\Omega})\vec{c}+\pa{\frac{-\vec{g}\cdot\vec{\Omega}}{\vec{g}\cdot(\vec{c}\times\vec{\Omega})}\int_0^{r^0}e^{-\phi(r)}dr+b(r^1)}\vec{\Omega}+(\vec{g}\cdot\vec{c})\vec{c}\times \vec{\Omega},
\end{gaeq*}%
with Riemann invariants
\begin{gaeq*}%\label{ex:GMC:25}
\vec{c}\cdot\vec{x}=\int_0^{r^0}e^{-\phi(r)}dr,\quad \vec{c}=(c_1,c_2,0),\\
\frac{\vec{c}\cdot\vec{x}}{\vec{g}\cdot(\vec{c}\times \vec{\Omega})} \br{a(r^1)-\vec{g}\cdot(\vec{c}\times\vec{\Omega})a_1(r^1)+b(r^1)a_2(r^1)+\vec{g}\cdot\vec{c}a_3(r^1)}\\
-\frac{1}{2}\vec{g}\cdot\vec{\Omega} a_2(r^1)\pa{\frac{\vec{c}\cdot \vec{x}}{\vec{g}\cdot(\vec{c}\times\vec{\Omega})}}^2+a(r^1)t+\bigg(a_1(r^1)\vec{c}\\
+a_2(r^1)\vec{\Omega}+a_3(r^1)\vec{c}\times\vec{\Omega}\bigg)\cdot\vec{x}=\psi(r^1),
\end{gaeq*}%
where $p_0>0$, $c_1$ and $c_2$ are arbitrary constants; $\rho(r^1)>0$, $\phi(r^0)$, $\psi(r^1)$, $b(r^1)$, $a(r^1)$ and $a_i(r^1)$ $(i=1,2,3)$ are arbitrary functions satisfying $\dot{b}a_2=0$, $\dot{\rho}\neq 0$, and $a_2\neq 0$ or $a_3\neq 0$ or $a\neq 0$.
\paragraph{\textbf{4}. The simple acoustic wave and the simple hydrodynamic state, $\mathfrak{AH}_0$.} In the case where $\vec{\lambda}^0\times\vec{\Omega}\neq 0$ and $\vec{\lambda}^0\cdot\vec{\Omega}=0$, $\del \vec{v}/\del r^1\cdot \vec{\Omega}=0$, $\del v/ \del r^1\cdot (\vec{\lambda}^0\times \vec{\Omega})=0$, $\vec{g}\cdot\vec{\Omega}=0$, $\kappa=3$ and $\del \vec{v}/\del r^{1}\cdot \vec{\lambda}^0\neq 0$, the solution is given by
\begin{gaeq*}%\label{ex:GMC:26}
\rho=\pa{\alpha(r^0,r^1)+\vec{g}\cdot(\vec{c}\times\vec{\Omega})}^{-1}S(r^1),\qquad p=a\rho^3,\\
\vec{v}=\alpha(r^0,r^1)\vec{c}+b_1\vec{\Omega}+\pa{\int_0^{r^0}e^{-\phi(r)}dr+T_1}\vec{c}\times \vec{\Omega},
\end{gaeq*}%
where $\alpha$ is given by the quadratic equation
\begin{gaeq}\label{ex:GMC:27}
\frac{1}{2}\pa{\alpha+\vec{g}\cdot (\vec{c}\times\vec{\Omega})}^2+\frac{3}{2}a(\alpha+\vec{g}\cdot(\vec{c}\times\vec{\Omega}))^{-2}S(r^1)^2+\epsilon \sqrt{3a}S(r^1)+S_1\\
-(\vec{g}\cdot\vec{c}-T_1)\int_0^{r^0}e^{-\phi(r)}dr+\frac{1}{2}\pa{\int_0^{r^0}e^{-\phi(r)}dr}^2=0,
\end{gaeq}%
and the Riemann invariants are given implicitly by
\begin{gaeq*}%\label{ex:GMC:28}
\vec{g}\cdot(\vec{c}\times\vec{\Omega})t+\vec{c}\cdot\vec{x}+c_1=\int_0^{r^0}e^{-\phi(r)}dr,\\
t+\epsilon(\sqrt{3a}\dot{S}(r^1))^{-1}\int_0^{r^0}e^{-\phi(r)}\frac{\del \alpha}{\del r^1}(r,r^1)dr=\psi(r^1).
\end{gaeq*}%
Here $c_1$, $a>0$, $b_1$, $T_1$ and $S_1$ are arbitrary constants, $\vec{c}=\vec{\lambda}^0/|\vec{\lambda}^0|$ is a constant vector such that $\vec{c}\cdot \vec{\Omega}=0$; $\epsilon=\pm 1$; $\phi(r^0)$, $S(r^1)\neq 0$, $\psi(r^1)$ and $\alpha(r^0,r^1)$ are any functions which satisfy (\ref{ex:GMC:27}) with $\dot{S}(r^1)\neq0$, ${\del \alpha}/{\del r^1}\neq 0$, $S(\alpha+\vec{g}\cdot(\vec{c}\times\vec{\Omega}))^{-1}>0$, and $\alpha\neq-\vec{g}\cdot(\vec{c}\times\vec{\Omega})$.
%######
\paragraph{}The generalized method of characteristics in the version presented in this paper proved to be a useful tool, since it led to many new interesting solutions. However, the superposition of simple waves and simple states is "linear" (according to the classification presented in \cite{Jeffrey:1976}) in the sense that the wave vector $\lambda^1$ does not change the direction of propagation on a simple state. So, it seems to be worthwhile to try to extend this method and check its effectiveness for the case when the Jacobian matrix $\del u^\alpha/\del x^i$ is not necessarily expressible in terms of simple integral elements of the form (\ref{eq:GMC:2}). This is, in short, the aim of the following sections.
%################## Multiwaves
\section{Multiwave solutions}\label{sec:2}
In this section we consider the possibility of generalizing the idea of Riemann wave solutions of quasilinear systems to what may be considered a nonlinear superposition of Riemann waves. The resulting solution will be called a Riemann $k$-wave or simply a $k$-wave. This has already been discussed for the simple wave in \cite{GrundlandLamothe:2013}. We suppose that the system is well-determined ($m=q$) and that it constitutes a system of $q$ equations and of $q$ dependent variables $u=(u^1,\ldots,u^q)$ in terms of $p$ independent variables $x=(x^1,\ldots, x^p)$. We look for a solution of system (\ref{eq:omr:1}) of the form
\begin{eqe}\label{eq:omr:2}
u=f(r^1(x,u),\ldots, r^k(x,u)),\qquad r^A(x,u)=\lambda^A_i(u)x^i,\quad A=1,\ldots, k,
\end{eqe}%
where the functions $r^A(x,u)$ are the Riemann invariants associated respectively with the real linearly independent wave vectors $\lambda^A(u)$ which satisfy the wave relation specified below. The Jacobian matrix $\del u$ takes the form
\begin{eqe}\label{eq:omr:3}
\del u=(u^\alpha_i)=\Phi^{-1}\frac{\del f}{\del r}\Lambda\in\RR^{q\times p},\qquad \Phi=I_q-\frac{\del f}{\del r}\frac{\del r}{\del u}\in\RR^{q\times q},
\end{eqe}%
where we use the notation
\begin{eqe}\label{eq:omr:4}
\frac{\del f}{\del r}=\pa{\frac{\del f^\alpha}{\del r^A}}\in\RR^{q\times k},\qquad
\Lambda=\pa{\lambda^A_i}\in\RR^{k\times p},\\
\frac{\del r}{\del u}=\pa{\frac{\del r^A}{\del u^\alpha}}=\frac{\del \Lambda_i}{\del u}x^i\in\RR^{k\times q},
\end{eqe}%
$r=(r^1,\ldots,r^k)$; $\alpha,\beta, \mu=1,\ldots,q$; $A=1,\ldots, k$; $i=1,\ldots,p$. Substituting (\ref{eq:omr:3}) into the system (\ref{eq:omr:1}) we obtain
\begin{eqe}\label{eq:omr:6}
\mathcal{A}^{\mu i}_\alpha\pa{\Phi^{-1}}^\alpha_\beta\frac{\del f^\beta}{\del r^A}\lambda^A_i=b^\mu
\end{eqe}%
or, in matrix form,
\begin{eqe}\label{eq:omr:7}
\pa{\mathcal{A}^i\lambda^A_i}\Phi^{-1}\frac{\del f}{\del r^A}=b,
\end{eqe}%
 where $\mathcal{A}^i=\pa{\mathcal{A}^{\mu i}_\alpha}\in\RR^{q\times q}$ and $b=(b^1,\ldots,b^q)^T$.
\paragraph{}Let us introduce $k$ real scalar functions $\Omega_A(x,u)$, $k$ matrices $L_A(x,u)\in \mathrm{SO}(q,\RR)$ and $k$ characteristic vectors $\tau_A(x,u)\in\RR^q$ such that
\begin{eqe}\label{eq:omr:8}
\Phi^{-1}\frac{\del f}{\del r^A}=\Omega_{(A)}L_Ab+\tau_A,\quad A=1,\ldots,k,
\end{eqe}%
\begin{eqe}\label{eq:omr:10}
\mathcal{A}^i\lambda_i^A\tau_A=0.
\end{eqe}%
Note that the vectors $\tau_A$ are characteristic vectors of the homogeneous part of equation (\ref{eq:omr:7}).
Next, we eliminate the quantities $\Phi^{-1}\pa{\del f/\del r^A}$ from equation (\ref{eq:omr:7}) through the relations (\ref{eq:omr:8}). We find an algebraic relation for the quantities $\Omega_A$, $\lambda^A$ and $L_A$
\begin{eqe}\label{eq:omr:9}
\sum_{A=1}^k\pa{\pa{\Omega_A\pa{\mathcal{A}^i\lambda_i^A}L_A}-I_q}b=0,
\end{eqe}%
 The relation (\ref{eq:omr:9}) represents a constraint on the form of the scalar functions $\Omega_A$, of the wave vectors $\lambda^A$ and of the rotation matrices $L_A$. We multiply equation (\ref{eq:omr:8}) on the left by the matrix $\Phi$ and then introduce the explicit expression for $\Phi$ given in (\ref{eq:omr:3}). Next, we solve the resulting equations for the matrices ${\del f}/{\del r}$ in order to obtain
\begin{eqe}\label{eq:omr:11}
\frac{\del f}{\del r}=\pa{\mathcal{L}_\beta b^\beta+\tau}\pa{I_k+\frac{\del r}{\del u}(\mathcal{L}_\beta b^\beta+\tau)}^{-1},
\end{eqe}%
where we have introduced the following notation
%\begin{eqe}\label{eq:omr:12}
$$
\tau=\pa{\tau_A^\alpha}\in\RR^{q\times k}\qquad\mathcal{L}_\beta=\pa{\Omega_{(A)}L^{\alpha}_{A\beta}}\in\RR^{q\times k}.
$$%\end{eqe}%
We assume that the $k\times k$ dimensional matrix $I_k+\pa{\del r/\del u}\pa{\mathcal{L}_\beta+\tau}$ is invertible. In order to determine the conditions for which the system (\ref{eq:omr:11}) is well defined in the sense that it represents a system for the functions $f^\alpha$ in terms of the invariants $r^A$ only, it is convenient to define a set of vector fields on the space $E\times\mathcal{U}$
\begin{eqe}\label{eq:omr:14}
X_a=\xi^i_a(u)\del_{x^i},\qquad a=1,\ldots,p-k,
\end{eqe}%
with the property
\begin{eqe}\label{eq:omr:15}
\xi^i_a\lambda_i ^A=0,\qquad a=1,\ldots,p-k,\quad A=1,\ldots,k.
\end{eqe}%
Applying the vector fields (\ref{eq:omr:14}) to the system (\ref{eq:omr:11}) and taking into account that $X_a$ annihilates all functions of the invariants $r^A$, we find the following constraints
\begin{eqe}\label{eq:omr:16}
X_a\pa{\pa{\mathcal{L}_\beta b^\beta+\tau}\pa{I_k+\frac{\del r}{\del u}(\mathcal{L}_\beta b^\beta+\tau)}^{-1}}=0, \qquad a=1,\ldots, p-k.
\end{eqe}%
In conclusion, if the conditions (\ref{eq:omr:10}), (\ref{eq:omr:9}) and (\ref{eq:omr:16}) are satisfied by an appropriate choice of the functions $\Omega_A$, $\lambda^A$, $\tau_A$ and $L_A$, then a solution of system (\ref{eq:omr:11}) represents a multiwave solution of the hyperbolic system (\ref{eq:omr:1}) of the proposed form (\ref{eq:omr:2}). It should be noted that these conditions are sufficient but not necessary.
\paragraph{}Let us compare the proposed approach with the GMC. If we suppose that there exists a solution of the form (\ref{eq:omr:1}) such that the matrix $\Phi$ is invertible, and that the vector $\tau_A$ obeys the wave relation (\ref{eq:omr:10}), then we can determine a function $\Omega_A$ and a rotation matrix $L_A$ in such a way that the relations (\ref{eq:omr:8}) are satisfied. Indeed, since the vectors $\Phi^{-1}(\del f/\del r^A)-\tau_A$ and $b$ have the same dimension, there exists a bijective transformation between these two vectors. This transformation is determined by the rotation matrix $L_A$ and the scalar function $\Omega_A$. Consequently, the Jacobian matrix of a solution takes the form
\begin{aleq}\label{eq:r:ast}
\frac{\del u}{\del x^i}&=\sum_A\pa{\Omega_A L^\alpha_{A\beta}+\tau_A^\alpha}\lambda_i^A\\
&=\sum_A\pa{\eta_A^\alpha(x,u)+\tau_A^\alpha(x,u)}\lambda_i^A(u),
\end{aleq}
where
%\begin{eqe}\label{eq:r:1}
$$
\eta_A^\alpha(x,u)=\Omega_{(A)}L^\alpha_{A\beta}b^\beta,\qquad \mathcal{A}^i\lambda_i^A\tau_A=0.
$$
%\end{eqe}%
So the Jacobian matrix of a solution admits the following decomposition
\begin{eqe}\label{eq:r:2}
\frac{\del u^\alpha}{\del x^{i}}=\zeta_1^\alpha\lambda^1_i+\cdots+\zeta^\alpha_{k}\lambda^k_i,
\end{eqe}%
or equivalently, (\ref{eq:r:2}) written in its matrix form, is
%\begin{eqe}\label{eq:r:3}
$$
\del u=\zeta_1\otimes \lambda^1+\cdots+\zeta_k\otimes\lambda^k,
$$%\end{eqe}%
where we have used the notation
%\begin{eqe}\label{eq:r:4}
$$
\zeta^\alpha_A(x,u)=\eta_A^\alpha(x,u)+\tau_A^\alpha(x,u).
$$
%\end{eqe}%
Note that, in general, conditions (\ref{eq:r:ast}) are weaker than the differential constraints (\ref{eq:GMC:2}) required by the GMC, since the latter are submitted to the wave relation (\ref{eq:GMC:ast}) and differential constraints (\ref{eq:GMC:3}) and (\ref{eq:GMC:5}). Indeed, (\ref{eq:omr:8}) implies, that all first-order derivatives of $u$ with respect to $x^i$ are decomposable in the form (\ref{eq:r:2}) on some open domain $\mathcal{B}\subset E\times \mathcal{U}$, where $\zeta=\pa{\zeta^\alpha_A(x,u)}$ are real-valued matrix functions defined on the first jet space $J^1=J^1(E\times \mathcal{U})$. This fact allows us to ease restrictions imposed on the initial data at $t=0$ for hyperbolic systems. Thus we are able to consider more diverse configurations of waves involved in superpositions (described by inhomogeneous systems) than in the GMC case.
\paragraph{}We end this section with two remarks:
\begin{itemize}
\item[  \textbf{i})] In the case where the matrices $\mathcal{L}_ \beta$ and $\tau$ depend only on the dependent variables $u$, conditions (\ref{eq:omr:16}) simplify to
%\begin{eqe}\label{eq:omr:17}
$$
X_a\pa{I_k+\frac{\del r}{\del u}(\mathcal{L}_\beta b^\beta+\tau)}^{-1}=0
$$
%\end{eqe}%
which is equivalent to
%\begin{eqe}\label{eq:omr:18}
$$
X_a\br{I_k+\frac{\del r}{\del u}(\mathcal{L}_\beta b^\beta+\tau)}=0.
$$%\end{eqe}%
Considering the notation (\ref{eq:omr:4}), we find the sufficient condition
%\begin{eqe}\label{eq:omr:19}
$$
\xi^i_a\frac{\del \Lambda_i}{\del u}\pa{\mathcal{L}_\beta b^\beta+\tau}=0,
$$%\end{eqe}%
for which the system (\ref{eq:omr:11}) is well defined in the sense that the right-hand side is expressed as a function of the invariants $r=(r^1,\ldots,r^k)^T$ only. More generally, a sufficient condition for the system (\ref{eq:omr:11}) to be expressible in terms of $r$ is that the matrix $(\del \Lambda_i/\del u)(\mathcal{L}_\beta b^\beta+\tau)$ be proportional to the matrix $\Lambda$. This means that there exists a scalar function $\sigma(u)$ such that
%\begin{eqe}\label{eq:omr:20}
$$
\pa{\frac{\del \Lambda_i}{\del u}\mathcal{L}_\beta b^\beta+\tau}=\sigma(u)\Lambda,
$$%\end{eqe}%
where $\Lambda$ is defined by (\ref{eq:omr:4}).
\item[ \textbf{ii})] If $k=p$, there are no orthogonality conditions (\ref{eq:omr:15}) and therefore the conditions (\ref{eq:omr:16}) are no longer necessary in order to ensure that the system (\ref{eq:omr:11}) is well defined. In this case however, the linear independence of the wave vectors $\lambda^A$ ensures that the relations $r^A=\lambda^A_i x^i$ can be inverted to obtain the $x^i$ in terms of the $r^A$. Consequently, the independent variables $x^i$ can be eliminated from the system (\ref{eq:omr:11}), which is then expressed in terms of $f$ and $r$ only.
\end{itemize}
%####################### modes multiples ####################################
\section{Multimode solutions}\label{sec:modemul}\label{sec:3}
The algebraization that has been performed in \cite{GrundlandLamothe:2013} for the elliptic system (\ref{eq:omr:1}) allows us to construct more general classes of solutions, namely the $k$-multimode solutions (a superposition of $k$ simple mode solutions). We look for real solutions of the form
\begin{gaeq}\label{eq:mm:1}
u=f(r^1(x,u),\ldots,r^k(x,u), \bar{r}^1(x,u),\ldots,\bar{r}^k(x,u)),\\
r^A(x,u)=\lambda_i^A(u)x^i,\quad \bar{r}^A(x,u)=\bar{\lambda}_i^A(u)x^i,\qquad A=1,\ldots,k.
\end{gaeq}%
The complex-valued wave vectors $\lambda^A$ and their complex conjugates $\bar{\lambda}^A$ are linearly independent. The Jacobian matrix $\del u$ of a real-valued solution (\ref{eq:mm:1}) in terms of Riemann invariants takes the form
\begin{eqe}\label{eq:mm:2}
\del u=\Phi^{-1}\pa{\frac{\del f}{\del r^A}\lambda^A+\frac{\del f}{\del\bar{r}^A}\bar{\lambda}^A}=\Phi^{-1}\pa{\frac{\del f}{\del r}\Lambda+\frac{\del f}{\del \bar{r}}\bar{\Lambda}},
\end{eqe}%
where
\begin{eqe}\label{eq:mm:4}
\Phi=I_q-\frac{\del f}{\del r}\frac{\del r}{\del u}-\frac{\del f}{\del \bar{r}}\frac{\del \bar{r}}{\del u}\in\RR^{q\times q},\quad \frac{\del f}{\del r}=\pa{\frac{\del f^\alpha}{\del r^A}}\in \CC^{q\times k},\quad \Lambda=\pa{\lambda_i^A}\in\CC^{k\times p},
\end{eqe}%
where $\del f/\del \bar{r}$ and $\bar{\Lambda}$ are complex conjugates of $\del f/\del r$ and $\Lambda$ respectively. We introduce the Jacobian matrix (\ref{eq:mm:2}) into the system (\ref{eq:omr:1}), which results in the following system written in matrix form
\begin{eqe}\label{eq:mm:5}
\mathcal{A}^i\pa{\lambda_i^A\phi^{-1}\frac{\del f}{\del r^A}+\bar{\lambda}_i^A\phi^{-1}\frac{\del f}{\del \bar{r}^A}}=b.
\end{eqe}%
We now define the complex-valued scalar and vector functions, $\Omega_A(x,u)$ and $\tau_A(x,u)$ respectively, as well as the matrices $L_A(x,u)\in \mathrm{SO}(q,\CC)$ such that
\begin{eqe}\label{eq:mm:6}
\Phi^{-1}\frac{\del f}{\del r^A}=\Omega_{(A)}L_Ab+\tau_A,\qquad \Phi^{-1}\frac{\del f}{\del \bar{r}^A}=\bar{\Omega}_{(A)}\bar{L}_A b+\bar{\tau}_A,\quad A=1,\ldots,k,
\end{eqe}%
where $\bar{\Omega}_A$, $\bar{\tau}_A$ and $\bar{L}_A$ are the complex conjugates of $\Omega_A$, $\tau_A$ and $L_A$ respectively. Substituting the expressions (\ref{eq:mm:6}) into the system (\ref{eq:mm:5}) and assuming that the characteristic vectors $\tau_A$ satisfy
\begin{eqe}\label{eq:mm:7}
\mathcal{A}^i\pa{\lambda_i^A\tau_A+\bar{\lambda}_i^A\bar{\tau}_A}=0,
\end{eqe}%
we obtain the algebraic condition
\begin{eqe}\label{eq:mm:8}
\pa{\pa{\sum_A\mathcal{A}^i\pa{\Omega_A\lambda_i^AL_A+\bar{\Omega}_A\bar{\lambda}_i^A\bar{L}_A}}-I_q}b=0,
\end{eqe}%
 on functions $\Omega_A$, $\lambda^A$ and $L_A$, $A=1,\ldots, k$ and their complex conjugates. Multiplying each of the equations (\ref{eq:mm:6}) on the left by $\Phi$ and replacing the explicit expression (\ref{eq:mm:4}) for the matrix $\Phi$, we obtain the following PDEs for the function $f$
\begin{aleq}\label{eq:mm:9}
\frac{\del f}{\del r^A}&=\pa{I_q-\frac{\del f}{\del r}\frac{\del r}{\del u}-\frac{\del f}{\del \bar{r}}\frac{\del \bar{r}}{\del u}}\pa{\Omega_{(A)}L_Ab+\tau_A},\\
\frac{\del f}{\del \bar{r}^A}&=\pa{I_q-\frac{\del f}{\del r}\frac{\del r}{\del u}-\frac{\del f}{\del \bar{r}}\frac{\del \bar{r}}{\del u}}\pa{\bar{\Omega}_{(A)}\bar{L}_Ab+\bar{\tau}_A}.\\
\end{aleq}%
Solving the system (\ref{eq:mm:9}) algebraically for the matrices $\del f/\del r$ and $\del f/\del \bar{r}$, we find the system in its canonical form
\begin{aleq}\label{eq:mm:10}
\frac{\del f}{\del r}&=\pa{I_q-\frac{\del f}{\del \bar{r}}\frac{\del \bar{r}}{\del u}}\pa{\mathcal{L}_\beta b^\beta+\tau}\pa{I_k+\frac{\del r}{\del u}\pa{\mathcal{L}_\beta b^\beta+\tau}}^{-1},\\
\frac{\del f}{\del \bar{r}}&=\pa{I_q-\frac{\del f}{\del r}\frac{\del r}{\del u}}\pa{\mathcal{L}_\beta b^\beta+\bar{\tau}}\pa{I_k+\frac{\del \bar{r}}{\del u}\pa{\bar{\mathcal{L}}_\beta b^\beta+\bar{\tau}}}^{-1},\\
\end{aleq}%
where we have used the notations
\begin{gaeq}\label{eq:mm:S}
\mathcal{L}_\beta=\pa{\Omega_{(A)}L^\alpha_{A\beta}}\in\CC^{q\times k},\qquad \bar{\mathcal{L}}_\beta=\pa{\bar{\Omega}_{(A)}\bar{L}^\alpha_{A\beta}}\in\CC^{q\times k}\\
S=\pa{\mathcal{L}_\beta b^\beta+\tau}\pa{I_k+\frac{\del r}{\del u}\pa{\mathcal{L}_\beta b^\beta+\tau}}^{-1}\in\CC^{q\times k},\qquad \tau=\pa{\tau^\alpha_A}\in\CC^{q\times k}
\end{gaeq}%
and $\bar{S}$ is its complex conjugate. Using (\ref{eq:mm:S}), we can write system (\ref{eq:mm:10}) in the form
\begin{aleq}\label{eq:mm:11}
\frac{\del f}{\del r}&=\pa{S- \bar{S}\frac{\del \bar{r}}{\del u}S}\pa{I_k+\frac{\del r}{\del u}\bar{S}\frac{\del \bar{r}}{\del u}S}^{-1},\\
\frac{\del f}{\del \bar{r}}&=\pa{\bar{S}- S\frac{\del r}{\del u}\bar{S}}\pa{I_k+\frac{\del\bar{r}}{\del u}S\frac{\del r}{\del u}\bar{S}}^{-1}.
\end{aleq}%
In order to verify that the system (\ref{eq:mm:11}) is well defined in the sense that the right-hand sides are expressible as functions of $r$ and $\bar{r}$ only, we apply the vector fields
\begin{eqe}\label{eq:mm:12}
X_a=\xi^i_a(u)\del_{x^i},\qquad i=1,\ldots,p,\quad a=1,\ldots,p-2k,
\end{eqe}%
to the relations (\ref{eq:mm:11}), where the coefficients $\xi^i_a(u)$ satisfy the orthogonality conditions
\begin{eqe}\label{eq:mm:13}
\xi^i_a(u)\lambda^A_i=0,\qquad \xi^i_a(u)\bar{\lambda}^A_i=0,\qquad A=1,\ldots,k.
\end{eqe}%
As a result, we obtain the conditions
\begin{aleq}\label{eq:mm:14}
X_a\br{\pa{S- \bar{S}\frac{\del \bar{r}}{\del u}S}\pa{I_k+\frac{\del r}{\del u}\bar{S}\frac{\del \bar{r}}{\del u}S}^{-1}}&=0,\qquad a=1,\ldots, p-2k,\\
X_a\br{\pa{\bar{S}- S\frac{\del r}{\del u}\bar{S}}\pa{I_k+\frac{\del \bar{r}}{\del u}S\frac{\del r}{\del u}\bar{S}}^{-1}}&=0,
\end{aleq}%
which are necessary and sufficient for the system (\ref{eq:mm:11}) to be well-defined in terms of the Riemann invariants $r$ and $\bar{r}$. In conclusion, if the conditions (\ref{eq:mm:7}),  (\ref{eq:mm:8}) and (\ref{eq:mm:14}) are satisfied, then integrating (\ref{eq:mm:11}), we obtain a solution describing a multimode solution of the system (\ref{eq:omr:1}).
%#################### multiwave for underdetermined system #####################################
\section{Multiwave solutions of underdetermined systems}\label{sec:5}
In Sections \ref{sec:2} and \ref{sec:modemul}, we established a method allowing the construction of solutions expressed in terms of Riemann invariants by restricting ourselves to well-determined systems ($q=m$). The proposed method can be modified so that it can be applied to underdetermined systems. Consider the first-order quasilinear hyperbolic system of PDEs (\ref{eq:omr:1}), where the number of equations $m<q$, and $q$ is the number of dependent variables. The expression for the Jacobian matrix given in equation (\ref{eq:omr:3}) remains valid with the matrices $\del f/\del r$, $\del r/\del u$ and $\Lambda$ defined in equation (\ref{eq:omr:4}). Therefore, by substituting the Jacobian matrix into the system (\ref{eq:omr:1}), we obtain equation (\ref{eq:omr:6}) for the case of underdetermined systems. In the well-determined case, the vectors $\Phi^{-1}\del f/\del r^A$ and $b$ were of equal dimensions, which allowed us to express the vectors $\Phi^{-1}\del f/\del r^A$ in terms of $b$ by the introduction of the scalar functions $\Omega_A(x,u)$ and the matrices $L_A(x,u)\in \mathrm{SO}(q,\RR)$. Since the system is now underdetermined, the vectors $\Phi^{-1}\del f/\del r^A$ can be expressed in terms of $b$ through certain rectangular matrices $P_A$ which are of dimension $q\times m$. The matrices $P_A$ are constructed from $m\times m$ matrices $L^\alpha_A(x,u)\in \mathrm{SO}(m,\RR)$, where $m$ is the number of equations in (\ref{eq:omr:1}), and are defined by
\begin{eqe}\label{eq:omrsd:1}
P_A=\Omega^\alpha_{(A)}(x,u)M_\alpha L^\alpha_A(x,u),\quad A=1,\ldots,k,
\end{eqe}%
where the $q\cdot m$ scalar functions $\Omega^\alpha_A$ are real-valued and the constant matrices $M_\alpha$ are defined by
%\begin{eqe}\label{eq:omrsd:2}
$$
M_\alpha=\pa{\delta_{\alpha i}\delta_{j1}}.
$$%\end{eqe}%
The vectors $\Phi^{-1}(\del f/\del r^A)$ can then be written as
\begin{eqe}\label{eq:omrsd:3}
\Phi^{-1}\frac{\del f}{\del r^A}=P_A+\tau_A,\quad A=1,\ldots,k,
\end{eqe}%
where the characteristic vectors $\tau_A\in\RR^q$ depend on the independent variables $x$ and the dependent variables $u$ and obey the relation
\begin{eqe}\label{eq:omrsd:4}
\mathcal{A}^i\lambda_i^A\tau_A,\quad A=1,\ldots,k,
\end{eqe}%
where $\mathcal{A}^i=\pa{A^{\mu i}_\alpha}\in \RR^{m\times q}$. Substituting the relations (\ref{eq:omrsd:3}) into equation (\ref{eq:omr:7}), we obtain the algebraic relation
\begin{eqe}\label{eq:omrsd:5}
\pa{\mathcal{A}^i\lambda_i^AP_A-I_m}b=0,
\end{eqe}%
where $I_m$ is the identity matrix of dimension $m\times m$. In order to obtain the reduced system of PDEs for the functions $f$ in terms of $r^A$, we proceed in a way similar to that for the multiwave solution presented in Section III. If we suppose that
\begin{eqe}\label{eq:omrsd:6}
\det\pa{I_k+\frac{\del r}{\del u}\pa{\mathfrak{p}_\mu b^\mu+\tau}}\neq 0,
\end{eqe}%
where $\mathfrak{p}_\mu=\pa{P_{A\mu}^\alpha}\in\RR^{q\times k}$, $\tau=\pa{\tau_A^\alpha}\in \RR^{q\times k}$, the reduced system takes the form
\begin{eqe}\label{eq:omrsd:7}
\frac{\del f}{\del r}=\pa{\mathfrak{p}_\mu b^\mu+\tau}\pa{I_k+\frac{\del r}{\del u}\pa{\mathfrak{p}_\mu b^\mu+\tau}}^{-1}.
\end{eqe}%
As in the well-determined case of Section \ref{sec:2}, the reduced system (\ref{eq:omrsd:7}) is well defined if the vector fields (\ref{eq:omr:14}), satisfying the orthogonality conditions (\ref{eq:omr:15}), cancel out the right-hand side of equation (\ref{eq:omrsd:7}),\ie
\begin{eqe}\label{eq:omrsd:8}
X_a\br{\pa{\mathfrak{p}_\mu b^\mu+\tau}\pa{I_k+\frac{\del r}{\del u}\pa{\mathfrak{p}_\mu b^\mu+\tau}}^{-1}}=0.
\end{eqe}%
In summary, if $k$ wave vectors $\lambda^A$ and $k$ matrices $P_A$, defined by (\ref{eq:omrsd:1}), satisfy the relation (\ref{eq:omrsd:5}), $k$ characteristic vectors $\tau_A$ satisfy the equation (\ref{eq:omrsd:4}), and if the condition (\ref{eq:omrsd:8}) is fulfilled, then the reduced system (\ref{eq:omrsd:7}) is well defined for $f$ in terms of the Riemann invariants $r^A$. Moreover, any solution of the system (\ref{eq:omrsd:7}) provides us with a multiwave solution of (\ref{eq:omr:1}) of the proposed form (\ref{eq:omr:2}).
%################ multimode for underdetermined system #########################
\section{Multimode solutions of underdetermined systems}\label{sec:6}
In analogy with the case of multiwaves presented above, we can consider underdetermined elliptic systems (\ref{eq:omr:1}) in order to obtain real multimode-type solutions in the form
\begin{gaeq}\label{eq:mmsd:1}
u=f(r^1(x,u),\ldots,r^k(x,u),\bar{r}^1(x,u),\ldots,\bar{r}^k(x,u)),\\
r^A(x,u)=\lambda^A_i x^i,\quad \bar{r}^A(x,u)=\bar{\lambda}^A_i x^i, \quad A=1,\ldots k,
\end{gaeq}%
where the $k$ linearly independent wave vectors $\lambda^A$ are complex-valued and $\bar{\lambda}^A$ are their complex conjugates. The Jacobian matrix takes the form (\ref{eq:mm:2}) with matrices $\Phi$, $\del f/ \del r$ and $\Lambda$ defined in equation (\ref{eq:mm:4}), and so relation (\ref{eq:mm:5}) is also satisfied in the underdetermined case. The difference between the present case and that of multiwaves presented above in Section \ref{sec:5} is that the functions $\Omega^\alpha_A(x,u)$ and the matrices $L^\alpha_A(x,u)\in \operatorname{SO}(m,\CC)$ are complex-valued functions. So, the relations (\ref{eq:omrsd:3}) have to be complemented by their complex conjugate relations
\begin{eqe}\label{eq:mmsd:2}
\Phi^{-1}\frac{\del f}{\del \bar{r}^A}=\bar{P}_A+\bar{\tau}_A,\quad A=1,\ldots,k,
\end{eqe}%
where $\bar{P}_A$ are the complex conjugate matrices of the $P_A$ defined in (\ref{eq:omrsd:1}) and the characteristic vectors $\bar{\tau}_A$ are complex conjugates of the vectors $\tau_A$ which, together, satisfy the relation
\begin{eqe}\label{eq:mmsd:3}
\mathcal{A}^i\pa{\lambda^A_i\tau_A+\bar{\lambda}^A_i\bar{\tau}_A}=0
\end{eqe}%
Making use of (\ref{eq:mmsd:2}) and eliminating $\Phi^{-1}{\del f}/{\del r}$ in equation (\ref{eq:mm:5}) we obtain the algebraic condition
\begin{eqe}\label{eq:mmsd:4}
\pa{\mathcal{A}^i\pa{\lambda^A_i P_A+\bar{\lambda}^A_i\bar{P}_A}+I_m}b=0.
\end{eqe}%
Hence, the system of PDEs for $f$ in terms of the $r^A$ and $\bar{r}^A$ takes the form
\begin{aleq}\label{eq:mmsd:5}
\frac{\del f}{\del r}=&\pa{I_q-\bar{\mathfrak{S}}\frac{\del \bar{r}}{\del u}}\mathfrak{S}\pa{I_k+\frac{\del r}{\del u}\bar{\mathfrak{S}}\frac{\del\bar{r}}{\del u}\mathfrak{S}}^{-1},\\
\frac{\del f}{\del \bar{r}}=&\pa{I_q-\mathfrak{S}\frac{\del r}{\del u}}\bar{\mathfrak{S}}\pa{I_k+\frac{\del \bar{r}}{\del u}\mathfrak{S}\frac{\del r}{\del u}\bar{\mathfrak{S}}}^{-1},\\
\end{aleq}%
where
$$\begin{aligned}
\mathfrak{S}=&\pa{\mathfrak{p}\mu b^\mu+\tau}\pa{I_k+\frac{\del r}{\del u}\pa{\mathfrak{p}_\mu b^\mu+\tau}}^{-1}\in\CC^{q\times k},\\
\mathfrak{S}=&\pa{\bar{\mathfrak{p}}\mu b^\mu+\bar{\tau}}\pa{I_k+\frac{\del \bar{r}}{\del u}\pa{\bar{\mathfrak{p}}_\mu b^\mu+\bar{\tau}}}^{-1}\in\CC^{q\times k},\\
\mathfrak{p}_\mu=&\pa{P^\alpha_{A\mu}}\in \CC^{q\times k}.
\end{aligned}$$%
By analogy with the previous cases in Section \ref{sec:5}, the system (\ref{eq:mmsd:5}) is well defined if the conditions
\begin{aleq}\label{eq:mmsd:6}
X_a\br{\pa{I_q-\bar{\mathfrak{S}}\frac{\del \bar{r}}{\del u}}\mathfrak{S}\pa{I_k+\frac{\del r}{\del u}\bar{\mathfrak{S}}\frac{\del\bar{r}}{\del u}\mathfrak{S}}^{-1}}=&0,\\
X_a\br{\pa{I_q-\mathfrak{S}\frac{\del r}{\del u}}\bar{\mathfrak{S}}\pa{I_k+\frac{\del \bar{r}}{\del u}\mathfrak{S}\frac{\del r}{\del u}\bar{\mathfrak{S}}}^{-1}}=&0,\\
\end{aleq}%
are satisfied, with the vector fields $X_a$ defined by (\ref{eq:mm:12}) and (\ref{eq:mm:13}).
\paragraph{}Thus we have demonstrated that, if $2k$ wave vectors $\lambda^A$ and $\bar{\lambda}^A$, and $2k$ matrices $P_A$ and $\bar{P}_A$, defined by (\ref{eq:omrsd:1}), satisfy the relation (\ref{eq:mmsd:4}), $2k$ characteristic vectors $\tau_A$ and $\bar{\tau}_A$ satisfy the equation (\ref{eq:mmsd:3}) and if the conditions (\ref{eq:mmsd:6}) are fulfilled, then the reduced system (\ref{eq:mmsd:5}) is well defined for $f$ in terms of the Riemann invariants $r^A$. Moreover, any solution of the system (\ref{eq:mmsd:5}) provides a multimode solution of the system (\ref{eq:omr:1}) of the proposed form (\ref{eq:mmsd:1}).
%################################ examples ##############
\section{Examples of appplications}\label{sec:7}
Now we illustrate the theoretical considerations presented in Sections \ref{sec:2}-\ref{sec:6} with various cases of inhomogeneous hydrodynamic-type systems.
\paragraph{\textbf{1.}}Consider a system of three equations for three unknown functions $u,v,w$ of the four independent variables $t,x,y,z$,
\begin{aleq}\label{ex1:eq:1}
u_t&+a_1 (u_x+v_y+w_z) + a_3 (u_z-w_x)-a_2 (v_x-u_y)=b_1(u,v,w),\\
v_t&+a_2 (u_x+v_y+w_z)+ a_3 (v_x-w_y)+a_1(v_x-u_y)=b_2(u,v,w),\\
w_t&+a_3 (u_x+v_y+w_z) - a_2 (v_z-w_y)- a_1(u_z-w_x)=b_3(u,v,w),
\end{aleq}%
where $U=(u,v,w)$ are the components of the velocity along the $x$, $y$, $z$, axes respectively and the $a_i\in \RR$, $i=1,2,3$, are real constants. In matrix form, the system (\ref{ex1:eq:1}) can be written as
%\begin{eqe}\label{ex1:eq:2}
$$
\frac{\del U}{\del t}+\mathcal{A}^1\frac{\del U}{\del x}+\mathcal{A}^2\frac{\del U}{\del y}+\mathcal{A}^3\frac{\del U}{\del z}=b(U),
$$%\end{eqe}%
where
\begin{gaeq*}%\label{ex1:eq:3}
U=(u,v,w)^T:\RR^4\rarrow \RR^3,\qquad b=(b_1,b_2,b_3)^T,\\
\mathcal{A}^1=\pa{\begin{array}{ccc}
           a_1 & -a_2 & -a_3 \\
           a_2 & a_1 & 0 \\
           a_3 & 0 & a_1
         \end{array}},\quad \mathcal{A}^2=\pa{\begin{array}{ccc}
           a_2 & a_1 & 0 \\
           -a_1 & a_2 & -a_3 \\
           0 & a_3 & a_2
         \end{array}},\quad \mathcal{A}^3=\pa{\begin{array}{ccc}
           a_3 & 0 & a_1 \\
           0 & a_3 & a_2 \\
           -a_1 & -a_2 & a_3
         \end{array}}.
\end{gaeq*}%
We consider the possibility of constructing mixed wave type solutions which may be nonlinear superpositions of a simple mode with a simple Riemann wave and we determine the admissible forms of the right-hand side $b$ of the system (\ref{ex1:eq:1}). This solution can be expressed in terms of a real-valued invariant $r_1$, of a complex-valued invariant $r_2$ and its complex conjugate $\bar{r}_2$. The invariants $r_1$, $r_2$ and $\bar{r}_2$ are associated with the real wave vector $\eta=(\eta_0,\eta_1,\eta_2,\eta_3)$, the complex-valued wave vector $\lambda=(\lambda_0,\lambda_1,\lambda_2,\lambda_3)$ and its complex conjugate $\bar{\lambda}$, respectively. So, we look for solutions of the form
 $$u=f(r_1,r_2,\bar{r}_2)=f(\eta x,\lambda x, \bar{\lambda} x).$$%
According to Section \ref{sec:2}, the wave vectors $\eta$, $\lambda$ and $\bar{\lambda}$ and the special orthogonal matrices $L_A$ have to satisfy the algebraic condition (\ref{eq:mm:8}), which, in the mixed case, takes the form
\begin{aleq}\label{ex1:eq:4}
\bigg(&\Omega_1 \pa{\eta_0 I_3+\eta_1 \mathcal{A}^1+\eta_2 \mathcal{A}^2+\eta_3 \mathcal{A}^3}L_1
+\Omega_2\bigg(\lambda_0 I_3+\lambda_1 \mathcal{A}^1+\lambda_2 \mathcal{A}^2+\lambda_3 \mathcal{A}^3\bigg)L_2\\
&+\bar{\Omega}_2\bigg(\bar{\lambda}_0 I_3+\bar{\lambda}_1 \mathcal{A}^1+\bar{\lambda}_2 \mathcal{A}^2+\bar{\lambda}_3 \mathcal{A}^3\bigg)\bar{L}_2-I_3\bigg)b=0,
\end{aleq}%
where $\Omega_1$ and the special orthogonal matrix $L_1$ are real-valued functions, but $\Omega_2$ and the special orthogonal matrix $L_2$ are complex-valued functions. The wave relation (which is here a mixed case of (\ref{eq:omr:10}) and (\ref{eq:mm:7})) for the real characteristic vector $\tau_1$ and the complex characteristic vector $\tau_2$ and its complex conjugate $\bar{\tau}_2$ can then be written as
\begin{aleq}\label{ex1:eq:5}
&\pa{\eta_0 I_3+\eta_1 \mathcal{A}^1+\eta_2 \mathcal{A}^2+\eta_3 \mathcal{A}^3}\tau_1+\pa{\lambda_0 I_3+\lambda_1 \mathcal{A}^1+\lambda_2 \mathcal{A}^2+\lambda_3 \mathcal{A}^3}\tau_2\\
&+\pa{\bar{\lambda}_0 I_3+\bar{\lambda}_1 \mathcal{A}^1+\bar{\lambda}_2 \mathcal{A}^2+\bar{\lambda}_3 \mathcal{A}^3}\bar{\tau}_2=0.
\end{aleq}%
The algebraic conditions (\ref{ex1:eq:4}) and (\ref{ex1:eq:5}) are satisfied if we choose
\begin{gaeq*}\label{ex1:eq:6}
\eta=\pa{-M,a_1,a_2,a_3},\quad M=a_1^2+a_2^2+a_3^2,\quad \lambda=\pa{1,i a_1,ia_2,ia_3},\quad i^2=-1,\\
 \Omega_2=\frac{(1-i M)b_3}{2b_1 \pa{M^2+1}},\qquad
L_2=\pa{\begin{array}{ccc}
                                       0 & -\sin(\theta) & -\cos(\theta) \\
                                       0 & \cos(\theta) & -\sin(\theta) \\
                                       1 & 0 & 0
                                     \end{array}},\\
\theta=\arctan\pa{\frac{b_2(b_1+b_3)}{b_1b_3-b_2^2}}+i\operatorname{arccosh}\pa{\frac{-\epsilon b_1(b_1^2+b_2^2)^{1/2}}{b_3(b_2^2+b_3^2)}}, \epsilon=\pm 1,
\end{gaeq*}%
where $\Omega_1$, $L_1\in \mathrm{SO}(3,\RR)$, $\tau_1$,  $\tau_2=(M+i)T$, with $T=(T_1,T_2,T_3)^T\in\RR^3$ are arbitrary functions of $x$ and $U$. The Riemann invariants associated with the wave vectors $\eta$, $\lambda$ and $\bar{\lambda}$ take the form
\begin{gaeq*}%\label{ex1:eq:7}
r_1=-M t+a_1 x+a_2 y+a_3 z,\\
r_2=t+i\pa{a_1 x+a_2 y+a_3 z},\quad \bar{r}_2=t-i\pa{a_1 x+a_2 y+a_3 z}.
\end{gaeq*}%
Since the quantities $a_i$ which appear in the system (\ref{ex1:eq:1}) are constant, the wave vectors $\eta$, $\lambda$ and $\bar{\lambda}$ are also constant. Therefore, the partial derivatives $\del \lambda_i/\del u^\alpha$ all vanish, and so the matrix $\Phi$ in expression (\ref{eq:mm:2}) is the identity, \ie $\Phi=I_3$. Consequently, the partial differential system (\ref{eq:mm:5}) in terms of invariants $r_1$, $r_2$ and $\bar{r}_2$ becomes
\begin{eqe}\label{ex:eq:8}
\frac{\del U}{\del r_1}=\Omega_1 L_1 b+\tau_1,
\end{eqe}%
\begin{aleq}\label{ex:eq:9}
\frac{\del U}{\del r_2}=&(1+i M)\left(\pa{\begin{array}{c}
                                         T_1 \\
                                         T_2 \\
                                         T_3
                                       \end{array}}+\pa{\frac{\epsilon \pa{b_1^2(b_1^2+b_2^2)-b_3^2(b_2^2+b_3^2)}^{1/2}}{(1+M^2)(b_1^2+b_2^2)^{1/2}}}\pa{\begin{array}{c}
                                         b_2/b_1 \\
                                         1 \\
                                         0
                                       \end{array}}\right.\\
                                       &\left.+\frac{i}{2(1+M^2)}\pa{\begin{array}{c}
                                         b_1 \\
                                         b_2 \\
                                         b_2
                                       \end{array}}\right),
\end{aleq}%
\begin{aleq}\label{ex:eq:10}
\frac{\del U}{\del \bar{r}_2}=&(1-i M)\left(\pa{\begin{array}{c}
                                         T_1 \\
                                         T_2 \\
                                         T_3
                                       \end{array}}+\pa{\frac{\epsilon \pa{b_1^2(b_1^2+b_2^2)-b_3^2(b_2^2+b_3^2)}^{1/2}}{(1+M^2)(b_1^2+b_2^2)^{1/2}}}\pa{\begin{array}{c}
                                         b_2/b_1 \\
                                         1 \\
                                         0
                                       \end{array}}\right.\\
                                       &\left.+\frac{-i}{2(1+M^2)}\pa{\begin{array}{c}
                                         b_1 \\
                                         b_2 \\
                                         b_2
                                       \end{array}}\right).
\end{aleq}%
Note that since the derivative $\del U/\del r_1$ is real and the functions $\Omega_1(x,u)$, $L_1(x,u)$ and $\tau_1(x,u)$ are real and arbitrary, they can be chosen so as to satisfy equations (\ref{ex:eq:8}). Instead of the system composed of (\ref{ex:eq:9}) and (\ref{ex:eq:10}), we consider a linear combination of these equations in order to obtain an equivalent system with the real quantities $(1-i M)(\del U/\del r_2)+(1+i M)(\del U/\del\bar{r}_2)$ and $i(1-iM)(\del U/\del r_2)-(1+i M)(\del U/\del\bar{r}_2)$ on the left hand side. The three real-valued functions $T_1$, $T_2$, $T_3$ appear in three of the six resulting equations. We choose these arbitrary functions in such a way that the three equations where they appear are satisfied. Thus we only have to solve the three remaining equations, namely
\begin{gaeq}\label{ex1:eq:11}
\frac{\del u}{\del \xi}=\frac{b_1}{1+M^2},\quad \frac{\del v}{\del \xi}=\frac{b_2}{1+M^2},\quad \frac{\del w}{\del \xi}=\frac{b_3}{1+M^2},\\
 \x=(1/2)((1-i M)r_2+(1+i M)\bar{r}_2)=t+M(a_1 x+a_2 y+a_3 z).
\end{gaeq}%
If, for a given inhomogeneous term $b=(b_1,b_2,b_3)$, the system (\ref{ex1:eq:11}) can be solved, then
$$U(r_1,\xi)=U(-M t+a_1 x+a_2 y+a_3 z,t+M(a_1 x+a_2 y+a_3 z))$$%
is a solution of system (\ref{ex1:eq:1}). For example, in the case where the function $b$ is defined by
$$b_1=-\frac{u}{a_1}(a_1^2-u^2),\quad b_2=-\frac{v}{a_2}(a_2^2-v^2),\quad b_3=-\frac{w}{a_3}(a_3^2-w^2),$$%
and $|u|<|a_1|$, $|v|<|a_2|$, and $|w|<|a_3|$, the solution of equation (\ref{ex1:eq:1}) takes the form
\begin{gaeq}\label{eq:solsech}
u=a_1\sech\pa{\frac{\xi}{1+M^2}+c_1(r_1)},\quad v=a_2 \sech\pa{\frac{\xi}{1+M^2}+c_2(r_1)},\\ w=a_3\sech\pa{\frac{\xi}{1+M^2}+c_3(r_1)},
\end{gaeq}%
where the $c_i(r_1)$, $i=1,2,3$, are arbitrary functions of one variable. This solution represents a solitonic bump-type wave. Moreover, if the vector $b$ of the inhomogeneous system (\ref{ex1:eq:1}) is defined by
\begin{gaeq*}
b_1=-(1-k_1^2(1-u_2))^{1/2}(1-u^2)^{1/2},\quad b_2=-(1-k_2^2(1-v_2))^{1/2}(1-v^2)^{1/2}, \\ b_3=-(1-k_3^2(1-w_2))^{1/2}(1-w^2)^{1/2},
\end{gaeq*}%
then the system (\ref{ex1:eq:1}) admits a cnoidal wave solution
$$u=\operatorname{cn}(\xi,k_1),\quad v=\operatorname{cn}(\xi,k_2),\quad w=\operatorname{cn}(\xi,k_3).$$%
Another possible solution is obtained when the vector $b$ of the inhomogeneous system (\ref{ex1:eq:1}) has the form
\begin{gaeq*}b_1=-(u^2-1)^{1/2}(u^2-k_1^2)^{1/2},\quad b_2=-(v^2-1)^{1/2}(v^2-k_2^2)^{1/2},\\ b_3=-(w^2-1)^{1/2}(w^2-k_3^2)^{1/2}.
\end{gaeq*}%
In that case, the obtained solution is given by
\begin{gaeq}\label{eq:solcn}
u=(1-\operatorname{cn}(\xi+c_1(r_1),k_1)^2)^{-1/2},\qquad v=(1-\operatorname{cn}(\xi+c_2(r_1),k_2)^2)^{-1/2},\\ w=(1-\operatorname{cn}(\xi+c_3(r_1),k_3)^2)^{-1/2}.
\end{gaeq}%
Equation (\ref{eq:solcn}) represents a bounded multisolitonic solution of the system (\ref{ex1:eq:1}) in terms of the Jacobi elliptic function $\operatorname{cn}$. The moduli $k_i$ of the elliptic function are chosen in such a way that $0<k_i^2<1$, $i=1,2,3$. This ensures that the elliptic solution possesses one real and one purely imaginary period and that for the real argument we have a real-valued solution.
%######################## example 2 ####################################
\paragraph{\textbf{2.}}Consider the hydrodynamic-type system with three dependent and four independent variables
\begin{eqe}\label{ex2:eq:1}
U_t+a\times\pa{\nabla \times U}=b,
\end{eqe}%
where $U=(u,v,w)^T$ is the velocity vector, which depends on $t,x,y,z$, $a=(a_1,a_2,a_3)$ is a constant vector and $b=\pa{\kappa a_2, - \kappa a_1, e^w}^T$, $\kappa\in \RR$. System (\ref{ex2:eq:1}) can be written in the matrix form
\begin{eqe}\label{ex:eq:2}
U_t+\mathcal{A}^1 U_x+\mathcal{A}^2 U_y+\mathcal{A}^3 U _z=b,
\end{eqe}%
where the matrices $\mathcal{A}^i$ are given by
%\begin{eqe}\label{ex2:eq:3}
$$
\mathcal{A}^1=\pa{\begin{array}{ccc}
          0 & a_2 & a_3 \\
          0 & -a_1 & 0 \\
          0 & 0 & -a_1
        \end{array}
},\quad \mathcal{A}^2=\pa{\begin{array}{ccc}
                   -a_2 & 0 & 0 \\
                   a_1 & 0 & a_3 \\
                   0 & 0 & -a_2
                 \end{array}
},\quad \mathcal{A}^3=\pa{\begin{array}{ccc}
                   -a_3 & 0 & 0 \\
                   0 & -a_3 & 0 \\
                   a_1 & a_2 & 0
                 \end{array}}.
$$%\end{eqe}%
As in the previous example, we look for a mixed-type solution, \ie of the form
$$u=f(r_1,r_2,\bar{r_2})=f(\eta x,\lambda x, \bar{\lambda} x),$$%
where the wave vector $\eta=\pa{\eta_0,\ldots,\eta_3}$ is real and the wave vectors $\lambda=(\lambda_0,\ldots,\lambda_3)$ and $\bar{\lambda}=\pa{\bar{\lambda}_0,\ldots, \bar{\lambda}_3}$ are complex conjugates. The scalar functions $\Omega_A$ and the special orthogonal matrices $L_A$ have to satisfy the algebraic relation (\ref{ex1:eq:4}), in which $\Omega_1$ and $L_1$ are associated with the real Riemann invariant $r_1$ and $\Omega_2$, $\bar{\Omega}_2$, $L_2$, $\bar{L}_2$ are associated with the complex Riemann invariants $r_2$ and $\bar{r}_2$. The real vector $\tau_1$ associated with $r_1$ and the complex-valued vectors $\tau_2$ and $\bar{\tau}_2$ associated with $r_2$ and $\bar{r}_2$ have to satisfy wave relation (\ref{ex1:eq:5}). The equations (\ref{ex1:eq:4}) and (\ref{ex1:eq:5}) are satisfied by the following choices
$$\eta=\pa{1,0,0,a_3^{-1}},\quad \lambda=\pa{0,1,i \mu, 0},\quad \mu\in \RR,\ i^2=-1,$$%
$$\Omega_1=\frac{e^w}{(\kappa^2a_1+e^{2w})^{1/2}\cos(\beta)+\kappa a_2 \sin(\beta)},$$%
$$\Omega_2=\frac{\mu a_2+i a_1}{2\mu a_2 a_3},\quad L_1=\pa{\begin{array}{ccc}
            \frac{|a_2|\cos(\beta)}{(a_1^2+a_2^2)^{1/2}} & \frac{\epsilon a_1 (e^w+a_2 \kappa \sin(\beta))}{Q} & \frac{a_2 e^w\sin(\beta)-\kappa a_1^2}{Q} \\
            \frac{-\epsilon a_1\cos(\beta)}{(a_1^2+a_2^2)^{1/2}} & \frac{\epsilon (a_2 e^w-\kappa a_1^2\sin(\beta))}{Q} & \frac{\epsilon a_1(\kappa a_2+e^w\sin(\beta))}{Q} \\
            \sin(\beta) & \frac{-\kappa a_1 \cos(\beta)}{(\kappa^2a_1^2+e^{2w})^{1/2}} & \frac{e^w\cos(\beta)}{(\kappa^2 a_1^2+e^{2w})^{1/2}}
          \end{array}
}$$%
$$L_2=\pa{\begin{array}{ccc}
            0 & \frac{\epsilon a_1 (e^w+\kappa a_2)}{Q} & \frac{\epsilon (a_1^2\kappa-a_2 e^w)}{Q} \\
            0 & \frac{\epsilon(a_2 e^{w}-\kappa a_1^2)}{Q} & \frac{\epsilon a_1(e^{w}+\kappa a_2)}{Q} \\
            1 & 0 & 0
          \end{array}},\quad \tau_1=\pa{\tau_{11},\tau_{12},-\frac{a_1\ta_11+a_2\tau_12}{a_3}}^T,$$%
$$\tau_2=\pa{s_1-i\pa{(a_2 s_2+a_3 s_3)(\mu a_2)^{-1}},s_2+i S_2,s_3 (1+i a_1 (\mu a_2)^{-1})}^T$$%
where $\beta$ is an arbitrary function of $x$ and $u$, $\epsilon$ is the sign of $a_2$ and $Q=(a_1^2+a_2^2)^{1/2}\pa{\kappa^2 a_1^2+e^{2w}}^{1/2}$ and $s_1$, $s_2$, $s_3$, $S_2$ are arbitrary functions of $u$. Consequently the conditions (\ref{ex1:eq:4}) and (\ref{ex1:eq:5}) are satisfied and we obtain the reduced system
\begin{gaeq}\label{ex2:eq:4}
\frac{\del u}{\del r_1}=\tau_{11}+\epsilon a_2 e^w G,\ \frac{\del v}{\del r_1}=\tau_{12}-\epsilon a_1e^w G,\ \frac{\del w}{\del r_1}=-\frac{a_1\tau_{11}+a_2\tau_{12}}{a_3}+e^w,\\
\frac{\del u}{\del r_2}=s_1-i\pa{\frac{a_2 s_2+a_3 s_3}{\mu a_2}}+\frac{\epsilon (e^{2w}+\kappa^2 a_1^2)^{1/2}}{2 \mu a_3 (a_1^2+a_2^2)^{1/2}}(\mu a_2+ i a_1),\\
\frac{\del v}{\del r_2}=s_2+i S_2-\frac{\epsilon (e^{2w}+\kappa^2 a_1^2)^{1/2}}{2 \mu a_3 (a_1^2+a_2^2)^{1/2}}(\mu a_2- i a_1),\\
\frac{\del w}{\del r_2}=s_3\pa{1+ \frac{i a_1}{\mu a_2}}+\frac{\kappa}{2\mu a_3}(\mu a_2+i a_1),
\end{gaeq}%
and the three conjugate equations for $\del u/\del \bar{r}_2$, $\del v/\del \bar{r}_2$ and $\del w/\del \bar{r}_2$.
\paragraph{}One particular solution of the system (\ref{ex2:eq:4}) takes the form
\begin{eqe}\label{ex2:eq:sol}\begin{aligned}
u(r_1,r_2,\bar{r_2})=&\frac{c_0}{2\mu}\pa{\frac{\del H(r_2,\bar{r}_2)}{\del r_2}+\frac{\del H(r_2,\bar{r}_2)}{\del \bar{r}_2}}+\frac{f_1(r_1)}{\mu}-\frac{i\kappa}{4} \frac{\eta-\xi}{\mu},\\
v(r_1,r_2,\bar{r_2})=&\frac{ic_0}{2}\pa{\frac{\del H(r_2,\bar{r}_2)}{\del r_2}-\frac{\del H(r_2,\bar{r}_2)}{\del \bar{r}_2}}+\frac{f_2(r_1)}{\mu}+\frac{\kappa}{4} \frac{\xi+\eta}{\mu},\\
w(r_1,r_2,\bar{r}_2)=&-a_3^{-1}\pa{\frac{a_1}{\mu}f_1(r_1)+a_2f_2(r_2)}\\
&-\ln\pa{-c_1-\int \exp\pa{-\frac{(\mu^{-1}a_1f_1(r_1)+a_2f_2(r_1))}{a_3}}dr_1},
\end{aligned}\end{eqe}%
where  $H$ is an arbitrary function of the complex Riemann invariants $r_2=x+i\mu y$ and $\bar{r}_2=x-i\mu y$, $\mu\in \RR$, $f_1$ and $f_2$ are arbitrary real functions of the real invariant $r_1=t+a_3^{-1}z$  and $c_0$, $c_1$ are real integration constants. Note that, the solution is real if and only if ${\del H}/{\del r_2}=\del \bar{H}/\del \bar{r}_2$ holds.
\paragraph{\bf{3.}}Now we shall consider an example which illustrates the theoretical considerations presented in Section \ref{sec:6}. We have chosen the underdetermined Loewner system \cite{Loewner:1954}
\begin{eqe}\label{ex3:eq:1}
u_y-v_x=h_1(u,v,\rho),\qquad (\rho u)_x+(\rho v)_y=h_2(u,v,\rho),
\end{eqe}%
where $h_1$ and $h_2$ are functions (to be determined) of the dependent variables, \textit{i.e.} the components of the velocity of the fluid $u$, $v$ and the density $\rho$. System (\ref{ex3:eq:1}) describes a stationary compressible planar fluid flow admitting a vorticity $h_1$ and some source of matter $h_2$. After the change of variables
%\begin{eqe}\label{ex3:eq:2}
$$
\rho=e^q,\qquad h_1=b_1,\qquad h_2=\rho b_1,
$$%\end{eqe}%
the system (\ref{ex3:eq:1}) becomes
%\begin{eqe}\label{ex3:eq:3}
$$
u_y-v_x=b_1,\qquad u_x+v_y+uq_x+vq_y=b_2,
$$%\end{eqe}%
or equivalently in matrix form
\begin{eqe}\label{ex3:eq:4}
\mathcal{A}^1U_x+\mathcal{A}^2U_y=b,
\end{eqe}%
where $b=(b_1,b_2)^T$, $U=(u,v,\rho)^T$ and
\begin{eqe}\label{ex3:eq:5}
\mathcal{A}^1=\pa{\begin{array}{ccc}
0 & -1 & 0\\
1 & 0 & u
\end{array}},\qquad \mathcal{A}^2=\pa{\begin{array}{ccc}
1 & 0 & 0\\
0 & 1 & v
\end{array}}.
\end{eqe}%
We are looking for a simple mode ($k$=1 in (\ref{eq:mmsd:1})), that is a $\rank$ 2 solution of (\ref{ex3:eq:4}) of the form
%\begin{eqe}\label{ex3:eq:7}
$$
u=f(r,\bar{r}),\quad r=x+i y,\quad \bar{r}=x-i y,
$$%\end{eqe}%
 associated with the constant wave vectors
%\begin{eqe}\label{ex3:eq:6}
$$
\lambda=(1,i),\qquad \bar{\lambda}=(1,-i),\\
$$%\end{eqe}%
where $i$ is the imaginary unit. The algebraic condition (\ref{eq:mm:8}), with matrices $\mathcal{A}^j$ given by (\ref{ex3:eq:5}), is satisfied for the matrix
%\begin{eqe}\label{ex3:eq:9}
$$
P=\pa{\begin{array}{cc}
-\frac{b_2}{2} \frac{-2b_1^2-b_1+i(b_1^2+b_2^2)+Q}{b_1^2+b_2^2} & \frac{1}{2}\frac{2b_1b_2+b_2^2+ib_1(b_1^2+b_2^3)+Q}{b_1^2+b_2^2}\\
\frac{1}{2}\frac{-b_2^2-2b_1b_2+ib_2(b_1^2+b_2)+Q}{b_1^2+b_2^2} & -\frac{1}{2}\frac{b_2(2b_2+1)-i(b_1^2+b_2^2)+Q}{b_1^2+b_2^2}\\
W\cos\theta(v+i u) & -W\sin\theta(v+iu)
\end{array}},
$$%\end{eqe}%
and its complex conjugate $\bar{P}$, where $W$ and $\theta$ are arbitrary functions of independent and dependent variables. The matrix $\Phi$ is equal to the identity matrix, since the wave vectors $\lambda$ and $\bar{\lambda}$ are constant vectors. Under these circumstances, the reduced system given by (\ref{eq:mmsd:5}) becomes
\begin{aleq}\label{ex3:eq:10}
\frac{\del u}{\del r}=&b_2\pa{1/2+(1+i)b_1}+\tau_1,\qquad \text{c.c.},\\
\frac{\del v}{\del r}=&b_1\pa{1/2}+(1-i)b_2+\tau_2,\qquad\text{c.c.},\\
\frac{\del \rho}{\del r}=&W\pa{b_1\cos\theta-b_2\sin\theta}(iu+v)+\tau_3,\qquad\text{c.c.},
\end{aleq}%
where each c.c. means the complex conjugate of the previous equation. The quantities $\tau_j$, $j=1,2,3$, are the components of the characteristic vector $\tau$ which satisfies the wave relation
%\begin{eqe}\label{ex3:eq:11}
$$
\mathcal{A}^j\pa{\lambda_j\tau+\bar{\lambda}_j\bar{\tau}}=0,\qquad \tau=(\tau_1,\tau_2,\tau_3)^T.
$$%\end{eqe}%
After solving the equation (\ref{eq:mmsd:3}) for the characteristic vector $\tau$, the system (\ref{ex3:eq:10}) reduces to the equations
\begin{gaeq}\label{ex3:eq:12a}
\bar{F}_{\bar{r}}-F_r=b_1(F,\bar{F},q),\\
F_r+\bar{F}_{\bar{r}}+Fq_r+\bar{F}q_{\bar{r}}=b_2(F,\bar{F},q),
\end{gaeq}%
where we have introduced the notation
%\begin{eqe}\label{ex3:eq:12b}
$$
F=u+ i v,\qquad \bar{F}=u- i v.
$$%\end{eqe}%
In particular, if  we choose the right-hand side of the equation (\ref{ex3:eq:4}) to be
%\begin{eqe} \label{ex3:eq:13}
$$
b_1=\kappa \ln\rho\pa{u^2+v^2},\qquad b_2=0,
$$%\end{eqe}%
then we can solve equations (\ref{ex3:eq:12a}) for the functions $F$ and $\bar{F}$
\begin{eqe}\label{ex3:eq:14}
F=\frac{2i\bar{f}'(\bar{r})}{\kappa |f|^2},\quad \bar{F}=\frac{-2 i f'(r)}{\kappa |f|^2},\quad q=|f|^2,
\end{eqe}%
where $f$ is an arbitrary function of $r=x+i y$ and $f'(r)=df/dr$. By virtue of (\ref{ex3:eq:14}) the simple mode solution of the system (\ref{ex3:eq:1}) is given by
\begin{aleq}\label{ex3:eq:15}
u(x,y)=&\frac{i}{\kappa |f(x+i y)|^4}\pa{\bar{f}'(x- iy)-f'(x+i y)},\\
v(x,y)=&\frac{1}{\kappa |f(x+i y)|^4}\pa{\bar{f}'(x- i y)+ f'(x+iy)},\\
\rho(x,y)=&\exp|f(x+i y)|^2.
\end{aleq}%
Here the rank-2 solution depends on one arbitrary function $f$ and its complex conjugate.
%################################## final remarks #######################################
\section{Final remarks}\label{sec:finalremarks}
We have presented a variety of new approaches to solving first-order quasilinear systems in terms of Riemann invariants. These methods proved to be particularly effective in delivering solutions which can be interpreted physically as superpositions between $k$ simple waves for hyperbolic systems or $k$ modes for elliptic ones. One of the interesting results of our analysis is the observation that the algebraization of these systems enables us to construct classes of solutions for which the matrix of the derivatives of unknown functions is expressible through special orthogonal matrices. This fact made it possible to obtain different types of solutions corresponding to different admissible choices of the special orthogonal matrices involved in this computation.
\paragraph{}Concerning the application to the inhomogeneous fluid dynamics equations (Section \ref{sec:ex:GMC}), we have shown that our approach is productive, leading to new interesting solutions. Using the GMC we were able to construct several $\rank 2$ solutions in the cases of $\mathfrak{EE}_0$, $\mathfrak{EA}_0$, $\mathfrak{EH}_0$, $\mathfrak{AE}_0$ and $\mathfrak{AH}$. Each of these solutions represents a superposition of a simple wave with a simple state and involves several arbitrary functions of one variable. The first derivatives of these solutions, in most cases, tend to infinity after a finite time. It was proved \cite{GrundlandVassiliou:1991,John:1974,Rozdestvenski:1983} that if the initial data is sufficiently small, then there exists a time interval, say $[t_0,T]$, for which the gradient catastrophe of a solution of the system (\ref{ex:GMC:1}) does not occur. For these types of solutions expressible in terms of Riemann invariants, we generalize the GMC in order to obtain wider classes of solutions of the inhomogeneous systems (see sections \ref{sec:3}-\ref{sec:7}). The proposed technique is applied to the inhomogeneous hydrodynamic-type systems (\ref{ex1:eq:1}), (\ref{ex2:eq:1}) and (\ref{ex3:eq:1}). We have introduced the complex integral elements instead of the real simple integral elements. This allows us to construct, based on those elements, several nonlinear superpositions of elementary solutions (modes) for these hydrodynamic-type systems, which to our knowledge are all new (namely (\ref{eq:solsech}), (\ref{eq:solcn}), (\ref{ex2:eq:sol}) and (\ref{ex3:eq:15})). This approach proves to be an effective tool for this purpose. The proposed technique is applicable to larger classes of hyperbolic and elliptic quasilinear systems. So it is worth investigating whether our approach to construct multimode solutions can be extended to the Navier--Stokes system in the presence of external forces (gravitational, Coriolis, etc.), where the dissipation effects are of particular interest. Some preliminary analysis suggests that this is feasible.
\paragraph{}It is worth noting that the method for solving first-order quasilinear differential equations can be applied, with necessary modifications, to more general cases, namely to nonautonomous systems for which the coefficients depend on unknown functions $u$ and also on independent variables $x$. An extension of our analysis to this case will be the subject of a future work.
%############ Acknowledgements #############
\section*{Acknowledgements}
This work was supported by a research grant from the Natural Sciences and Engineering Council of
Canada.
\section{Appendix: The simple state solutions \cite{GrundlandZelazny:1983}}
A mapping $u:\RR^p\rarrow \RR^q$ is called a simple state solution of the inhomogeneous system (\ref{eq:omr:1}) if all first-order derivatives of $u$ with respect to $x^i$ are decomposable in the following way
\begin{eqe}\label{eq:A:1}
du^\alpha(x)=\frac{\del u^\alpha}{\del x^i}dx^i=\gamma_0^\alpha(u)\lambda_i^0(u)dx^i,
\end{eqe}%
where the real-valued functions $\lambda^0=\pa{\lambda_1^0,\ldots,\lambda_p^0}\in E^\ast$ and $\gamma_0=(\gamma_0^1,\ldots,\gamma_0^q)\in T_u\mathcal{U}$ satisfy the algebraic relation
%\begin{eqe}\label{eq:A:2}
$$
\mathcal{A}^{\mu i}_\alpha(u)\lambda_i^0\gamma_{0}^\alpha=b^\mu(u).
$$%\end{eqe}%
In contrast to the condition
$$du^\alpha(x)=\xi\gamma^\alpha\lambda_idx^i,\quad \mathcal{A}_\alpha^{\mu i}(u)\lambda_i\gamma^\alpha=0,$$%
defining the simple wave solution for homogeneous systems, the expression (\ref{eq:A:1}) does not include a function $\xi$ of $x$. Consequently, the compatibility conditions are not identically satisfied and they lead to the following conditions
\begin{eqe}\label{eq:A:totDer}
d(du)=d\gamma_0\wedge \lambda^0+\gamma_0\otimes d\lambda^0=0,
\end{eqe}%
whenever equations (\ref{eq:A:1}) hold, where
$$\begin{aligned}
d\gamma_0=&\pa{\frac{\del \gamma_0}{\del u^\alpha}}du^\alpha=\gamma_{0,\gamma_0}\otimes \lambda^0,\quad \lambda^0=\lambda^0_i(u)dx^i,\\
d\lambda^0=&du^\alpha\wedge\frac{\del \lambda^0}{\del u^\alpha}=\lambda^0\wedge\lambda^0_{,\gamma_0}.
\end{aligned}$$%
Here we have used the notation $x_{,y}=\frac{\del x}{\del u^\alpha}y^\alpha$. Hence the system (\ref{eq:A:1}) has a solution if
$$\lambda^0_{,\gamma_0}\wedge \lambda^0=0$$%
holds. This means that the direction of $\lambda^0$ is constant along the vector field $\gamma_0$. Chosing a proper normalization for the wave vector $\lambda^0\rarrow a\lambda^0$ and the vector field $\gamma_0\rarrow a^{-1}\gamma_0$, where $a=a(u)$, one can obtain that $\lambda_0$ is constant along the vector field $\gamma_0$, \textit{i.e.}
$$\lambda_{0,\gamma_0}=0.$$%
Thus the image of a solution is a curve $u=f(r^0)$ tangent to the vector field $\gamma_0$. So one can choose a parametrization of a solution $u=f(r^0)$ such that the ordinary differential equations
$$\frac{df^\alpha}{dr^0}=\gamma_0^\alpha(f(r^0))$$%
hold. The wave vector $\lambda^0$ has a constant direction and by choosing a proper length of $\lambda^0$ such that $\del \lambda^0/\del r^0=0$, we may represent a simple state solution of the inhomogeneous system (\ref{eq:omr:1}) in the form
%\begin{eqe}\label{eq:A:125}
$$
u=f(r^0),\quad r^0=\lambda^0_i x^i,
$$%\end{eqe}%
where
$$\mathcal{A}_\alpha^{\mu i}(u) \lambda_i\gamma_0^\alpha=b^\mu(u),\quad \frac{\del f^\alpha}{\del r^0}=\gamma_0^\alpha(f(r^0)).$$%
This solution was introduced in analogy with a simple wave solution of a homogeneous system (\ref{eq:omr:1}) which satisfies the relations (see \textit{e.g.} \cite{Jeffrey:1976,Peradzynski:1985,Rozdestvenski:1983}).
\begin{gaeq*}%\label{eq:A:5}
u=f(r),\qquad r=\lambda_i(f(r))x^i,\\
\mathcal{A}^{\mu i}_\alpha(u)\lambda^i\gamma^\alpha=0,\qquad \frac{\del f^\alpha}{\del r}=\gamma^\alpha(f(r)).
\end{gaeq*}%
\bibliographystyle{alpha}% plain,unsrt,alpha

\begin{thebibliography}{10}
\bibitem{GrundlandVassiliou:1991}
Grundland~A.M., Vassiliou~P.:
\newblock On the solvability of the Cauchy problem for the Riemann double waves by the Monge--Darboux method.
\newblock {Analysis 11} 221-278, (1991)
\end{thebibliography}

\end{document}